\documentclass[runningheads]{llncs}
\usepackage{graphicx}
\usepackage{mathtools}
\usepackage{amsfonts}
\usepackage{amssymb,amsmath}
\usepackage{tikz}
\usepackage{algorithm}
\usepackage{algpseudocode}
\usepackage{bigints}
\DeclareMathAlphabet{\pazocal}{OMS}{zplm}{m}{n}

\begin{document}
\title{Externalities in Chore Division}
\titlerunning{Externalities in Chore Division}
%
\author{Mohammad Azharuddin Sanpui\orcidID{0000-0001-5030-9645}}
\authorrunning{Mohammad Azharuddin Sanpui }
%
\institute{Indian Institute of Technology Kharagpur, Kharagpur - 721302, India }

\maketitle               
\begin{abstract}
The chore division problem simulates the fair division of a heterogeneous, undesirable resource among several agents. In the fair division of chores, each agent only gets the disutility from its own piece. Agents may, however, also be concerned with the pieces given to other agents; these externalities naturally appear in fair division situations. We first demonstrate the generalization of the classical concepts of proportionality and envy-freeness while extending the classical model by taking externalities into account. Our results clarify the relationship between these expanded fairness conceptions, evaluate whether fair allocations exist, and examine whether fairness in the face of externalities is computationally feasible. In addition, we identify the most efficient allocations among all the allocations that are proportional and swap envy-free. We present tractable methods to achieve this under different assumptions about agents’ disutilities.

\keywords{Chore Division \and Fair Division \and Game Theory}
\end{abstract}
\section{Introduction}
The problem of allocating a heterogeneous, divisible resource among a set of $n$ agents with varying preferences is essentially described as the problem of cutting a cake. The cake cutting problem is a fundamental topic in the theory of fair division  \cite{brams1996fair,Cake,moulin2004fair,brandt2016handbook,ANoteOnCakeCutting} and it has received a significant amount of attention in the domains of mathematics, economics, political science, and computer science \cite{TheEfficiencyoffairdivision,cakecuttingreallyisnotapieceofcake,feeiciencyoffairdivisionwithconnectedpiece,ChildrenCryingatbirthdayparties,cakecuttingnotjustchildplay,TheQuerycomplexityofcakecutting}. Dividing a cake fairly among agents is a challenging task.
\textit{Envy-freeness} and \textit{proportionality} are the most important criteria of a fair allocation in the cake-cutting literature. In an envy-free allocation, every agent is pleased with the pieces they are allocated as opposed to any other agent's allocation \cite{ANENVY-FREECAKEDIVISIONPROTOCOL,ADiscreteandBoundedEnvy-FreeCakeCuttingProtocolforAnyNumberofAgents}. In a proportional allocation, each agent receives at least $1/n$ of the value he estimates to the cake \cite{Howtocutacakefairly}. When all of the cake has been divided, envy-freeness entails proportionality.

The two concepts of fairness are obviously fundamentally distinct upon closer examination. The concept of envy-freeness itself assumes that agents compare their own allocations to those of others, while proportionality just needs each agent to assess the quality of their own allocation in comparison to their best possible allocation.
This first assumption, which comes from psychological research, argues that agents are influenced both by their own piece of cake and the pieces of other agents. This type of influence is classified as ``externalities". These externalities may be either positive or negative.

There are several reasons for including externalities into consideration in an allocation problem. Externalities are usually prices or extra advantages that aren't passed on to customers through pricing and may be paid for by someone outside of the transaction. A scenario like this exists often, primarily while allocating resources. When agents share resources, an agent may be able to use resources given to a friend or family member because the agent has the right to use the resource. There are several merit commodities that provide positive demand externalities. In the medical field, vaccinations have positive externalities since they reduce the danger of infection for others around them. An example of a negative externality is the fact that when scheduling advertising time slots, Coca-Cola hurts if Pepsi gets the better slots.

Velez \cite{Velez2016FairnessAE} first addresses the study of externalities in fair division of indivisible goods and extends the notion of envy-freeness: \textit{swap envy-freeness} in which an agent cannot enhance by swapping its allocation with that of another agent. Branzei et al. \cite{Brnzei2013ExternalitiesIC} initiate the idea of externalities in fair division of heterogeneous divisible resources (or cakes) and extend the fairness notions: proportionality and swap stability. Swap stability requires that no agent can benefit by swapping the allocations between any pair of agents. Proportionality requires that each agent get the value of at least $1 \slash n$ in the comparison with her best possible allocation.

In contrast, the dual problem of cake-cutting, also known as chore division, seeks to allocate an undesirable resource to a set of $n$ agents, with each agent wishing to receive as little of the resource as possible. Chore division might model the allocation of chores within a household, liabilities in a bankrupt company, etc. Similar to the cake-cutting problem, dividing a chore fairly is also a challenging problem. The most important criteria of a fair allocation, similar to cake cutting, are \textit{envy-freeness} and \textit{proportionality}. Though several algorithms for cake cutting also apply to chore division, the theoretical properties of the two problems differ in many cases, and much less work has been done for chore division than for cake cutting \cite{AlgorithmicSolutionsforEnvy-FreeCakeCutting,AlgorithmsforCdivisionofchores,complexitychoredivisio,DividingConnectedChoresfairly,peterson1998exact,Chaudhury2020DividingBI}.

The problem of chore division with externalities, commonly referred to as ``negative externalities," is the inverse of the problem of cake-cutting with externalities, commonly referred to as ``positive externalities." In the presence of positive externalities, agents get benefits from the allocations assigned to other agents, but in the case of negative externalities, agents have to pay costs associated with the allocations gained by other agents.
In economics, a negative externality is when one party pays a price due to another party's activities. When one party, such as a corporation, makes another party less fortunate while avoiding paying the expenses, this is a negative externality. In environmental economics, negative externalities are essential when pollution represents a significant cost incurred by outside parties.  In various settings for resource allocation, negative externalities may also occur. For instance, when allocating resources to opposing groups, allocating a crucial resource to another group could make it more difficult for one group to flourish.
Our aim throughout the entire paper is to study chore division with externalities using the extended model of cake-cutting with externalities presented by Branzei et al. \cite{Brnzei2013ExternalitiesIC}.
\subsection{Previous Work: Externalities in Cake Cutting}
Positive externalities, broadly defined, are benefits that do not pass through prices to a party outside of the transaction. As an example, vaccination not only confers protection against illness upon the individual who receives it but also extends that protection to all those in the vicinity. Positive externalities that occur during the adoption of new technologies are referred to as network effects in games of network formation \cite{easley2010networks}. For instance, with the introduction of the phone, the perceived worth of the device to a prospective buyer is contingent upon the number of individuals who concurrently use a phone. Positive externalities are significant in resource allocation contexts, as the allocation of resources by one individual might have an influence on others. These circumstances are especially significant in the setting of social networks, since agents benefit from the allocations of others because of the presence of interactions. Consider the case where each agent is attempting to execute an online project and is assigned certain time slots on a server. During the downtime of their colleagues, the agents might have the opportunity to conduct more experiments and enhance the project's quality. Similarly, agents equipped with the best technology may utilize land (such as for road building or agricultural harvesting) more effectively, and everyone can profit from their efforts.

Using the expanded model of cake-cutting with externalities proposed by Branzei et al.\cite{Brnzei2013ExternalitiesIC}, our goal is to explore chore division with externalities across the whole article. They discuss many results in the context of cake-cutting while taking externalities into account. Some results are easily applicable to chores. In the existence of swap stability under weak assumptions, they provide an upper limit on the number of cuts required. This result is valid for the existence of swap-stable allocation for chore division with externalities (Theorem $3$). They demonstrate that, even with two agents, the generalized Robertson-Webb communication model cannot provide a proportional protocol. This assumption remains true even when chores have externalities (Theorem $6$).
\subsection{Related Works}
Externality theory is extensively researched in economics \cite{Ayres1969ProductionCA,Katz1985NetworkEC}, but recently it has also been getting more attention from researchers of computer science \cite{Alon2012SequentialVW,Krysta2010CombinatorialAW,Haghpanah2011OptimalAW,Michalak2009OnRC,Brnzei2013MatchingsWE}. Velez \cite{Velez2016FairnessAE} addresses externalities in the fair division of indivisible items and money. He independently proposes swap envy-freeness, where no agent may improve their allocation by swapping it with another. Branzei et al. \cite{Brnzei2013ExternalitiesIC} first present a generalized cake-cutting model with externalities. They introduce the idea of swap stability, which is a more robust version of swap envy-freeness, and assert that no agent can profit from a swap involving any two agents. Several AI articles explore the fair division of indivisible items with externalities \cite{ghodsi2018fair,mishra2022fair,seddighin2019externalities,aziz2021fairness}.

In recent times, there has been an increasing focus on the issue of social welfare in the context of cake-cutting. The studies of this specific topic started by Caragiannis et al.~\cite{caragiannis2012efficiency}, with the objective of quantifying the demise in social welfare that may potentially arise from various fairness criteria.
Aumann et al.~\cite{aumann2012computing} investigated the problem of finding a contiguous allocation that maximizes utilitarian social welfare for both divisible and indivisible items.
Bei et al.~\cite{bei2012optimal} and Cohler et al.~\cite{Cohler2011OptimalEC} both looked at how to maximize utilitarian social welfare while adding extra fairness constraints of proportionality and envy-freeness.
Bei et al.~\cite{bei2012optimal} also developed approximation results for maximizing utilitarian social welfare with proportionality as a constraint. In this paper, we especially focus on minimizing utilitarian social welfare with extended fairness as a constraint for the chore division with externalities. Several AI articles explore the efficient allocations of cake \cite{Bei2012OptimalPC,Aumann2012ComputingSC,Arunachaleswaran2019FairAE}.

\subsection{Our Results}
Throughout the entire paper, we focus on the analysis of the chore division problem with externalities. This is the mirror image of the cake-cutting problem with externalities. Our focus is on three extended fairness notions: \textit{proportionality}, \textit{swap envy-freeness}, and \textit{swap stability}. Proportionality among $n$ agents requires that each agent needs to obtain at most $1/n$ of the disutility she gains based on the worst possible allocation from her point of view. Swap envy-freeness requires that no agent can make their allocation better by swapping it with another agent’s. Swap stability requires that no agent can benefit from swapping the allocations between any two agents. In Section $3$, we demonstrate the relationship between fairness properties. In Section $4$, we display the lower and upper bounds of the required number of cuts to get various fair allocations. In Section $5$, we affirm that the generalized Robertson-Webb communication model cannot provide a finite proportional protocol even for two agents. In Section $6$, we obtain the most efficient proportional and swap envy-free allocations under restricted disutility functions. We describe tractable methods for just doing so when the agents' disutilities are piecewise constant functions. We also deal with general disutility functions and an arbitrary number of agents. A method is designed to calculate optimal extended fair allocations by approximating disutility functions using piecewise constant functions. This method has a polynomial time-based complexity of ${1}/{\epsilon}$, where $\epsilon$ represents the deviation from optimality and the permissible level of envy.
\section{Preliminaries}
In intuitive terms, the challenge of allocating a heterogeneous, divisible resource among $n$ agents with each agent influencing other agents' allocations is known as the ``problem of cake-cutting with externalities." Branzei et al. \cite{Brnzei2013ExternalitiesIC} first proposed the general model for cake cutting with externalities. On the other hand, the dual problem of cake-cutting with externalities, also known as chore division with externalities, tries to divide an unwanted resource among n agents in a way that takes into account how each agent’s allocation is affected by the allocations of the other agents.

We consider a setting where a heterogeneous undesirable resource (or chore) represented by the interval $[0,1]$ is to be allocated among a set $N=[n]=\{1,2,\ldots,n\}$ of agents under externalities. Each agent $i$ is endowed with $n$ integrable, non-negative disutility functions $v_{i,j}:[0,1]\rightarrow \mathbb{R}_{\geq 0}$ which measures the disutility of agent $i$ for different parts of the chore allocated to agent $j$. A piece of chore $X$ is a countable union of disjoint subintervals of $[0,1]$. The disutility that agent $i$ obtains from a piece $X$ that is assigned to agent $j$ is $V_{i,j}(X)=\int_{x\in X}v_{i,j}(x)dx$. Defined in this manner, agent disutilities are additive, i.e., $V_{i,j}(X\cup Y ) = V_{i,j}(X) + V_{i,j}(Y)$ if $X$ and $Y$ are disjoint, and non-atomic, i.e., $V_{i,j}([x,x]) = 0$. Because of the latter property, we can treat open and closed intervals as
equivalent. In the classical model of chore division, $V_{i,j}(X)=0$ for all pieces $X\subseteq [0,1]$ and $i \neq j$. 

An allocation $A = (A_1,A_2,\ldots,A_n)$ is an assignment of a piece of chore $A_i$ to each agent $i$, such that the pieces are disjoint and $\bigcup_{i\in N} A_i\subseteq [0,1]$. Moreover, each piece $A_i$ is a possibly infinite set of disjoint subintervals of $[0,1]$. The  disutility of agent $i$ under allocation $A$ is $V_i(A)=\sum_{j=1}^n V_{i,j}(A_j)$.
  
Similarly to the classical model, disutilities are normalized so that all the agents have equal disutility. That is, the sum of each agent's multiple disutilities should be normalized to $1$ i.e., $\sum_{j=1}^n V_{i,j}([0,1])=1$ for all $i$. In the classical model of chore-division, $V_{i,j}([0,1])=0$ for all $i\neq j$ and then $V_{i,i}([0,1])=1$ for all $i$, this represents the chore division problem. Moreover, it may happen that for each agent $i$, $V_{i}(\Tilde{A}_i)=1$, where $\Tilde{A}_i$ is the worst possible allocation for agent $i$.\newline
\textbf{Example 1.} Suppose there are $n$ agents. Consider the disutility functions: $v_{i,i}(x)=1, \forall x\in [\frac{i-1}{n},\frac{i}{n}]$ and $v_{i,j}(x)=1, \forall x\in [\frac{j-1}{n},\frac{j}{n}]$ and for all $i\neq j$. Then $A=(A_1,A_2,\ldots,A_n)$ where $A_i=[\frac{i-1}{n},\frac{i}{n}]$ for all $i$, is the allocation such that $V_i(A)=1$. Thus, $A$ is the worst allocation for every agent $i$.

\subsection{Fairness notions}
\textit{Envy-freeness} and \textit{proportionality} are the most important fairness criteria in the theory of fair division. Branzei et al. \cite{Brnzei2013ExternalitiesIC} extend the definition of proportionality. 
\\
\textbf{Definition 1 (Proportionality).} An allocation $A=(A_1,A_2,\dots,A_n)$ is said to be \textit{proportional} if for every agent $i\in N$, $V_i(A)\leq \frac{1}{n}$ i.e., $\sum_{j=1}^n V_{i,j}(A_j)\leq \frac{1}{n}$.

{In words, each agent obtains at most $1/n$ of the disutility she receives based on the worst possible allocation from her point of view. This definition addresses and guarantees the classical definition: when there are no externalities, each agent receives at most her average disutility of the entire chore.}

Velez \cite{Velez2016FairnessAE} defines the notion of swap envy-freeness in the fair division of indivisible goods and money with externalities, in which each agent has no interest in exchanging her allocation with that of another agent. 
\\
\textbf{Definition 2 (Swap Envy-freeness).} An allocation $A=(A_1,A_2,\ldots,A_n)$ is \textit{swap envy-free} if for any two players $i,j\in N $, $V_{i,i}(A_i)+V_{i,j}(A_j)\leq V_{i,i}(A_j)+V_{i,j}(A_i)$.

{In other words, no agent can make their allocation better by swapping it with another agent's. This definition directly addresses and indicates the classical definition of envy-freeness when there are no externalities. }

Branzei et al. \cite{Brnzei2013ExternalitiesIC} introduce swap stability, which is a stronger version of swap envy-freeness in which no agent can benefit by swapping the allocations between any two agents.
\\
\textbf{Definition 3 (Swap Stability).} An allocation $A=(A_1,A_2,\ldots,A_n)$ is \textit{swap stable} if for any three players $i,j,k\in N $, $V_{i,j}(A_j)+V_{i,k}(A_k)\leq V_{i,j}(A_k)+V_{i,k}(A_j)$.

{Note that the definition of swap stability implies swap envy-freeness.}

For any $\epsilon>0$, the definitions of $\epsilon$-proportionality and $\epsilon$-swap envy-freeness are shown below.
\newline
\textbf{Definition 4 ($\epsilon$-Proportionality).} An allocation $A=(A_1,A_2,\dots,A_n)$ is said to be $\epsilon$-\textit{proportional} if for every agent $i\in N$, $V_i(A)\leq \frac{1}{n}+\epsilon$ i.e., $\sum_{j=1}^n V_{i,j}(A_j)\leq \frac{1}{n}+\epsilon$.
\newline
\textbf{Definition 5 ($\epsilon$-Swap envy-freeness).} An allocation 
$A=(A_1,A_2,\ldots,A_n)$ is $\epsilon$-\textit{swap envy-free} if for any two players $i,j\in N $, $V_{i,i}(A_i)+V_{i,j}(A_j)\leq V_{i,i}(A_j)+V_{i,j}(A_i)+\epsilon$.
\subsection{Social welfare}
Efficiency, often known as social welfare, is a significant measure for evaluating an allocation. Efficiency, compared to fairness, evaluates the solution's performance from a global point of view. The efficiency of an allocation may be precisely defined as the sum of all agents' disutilities concerning the allocation. Efficiency is, therefore, a concrete quantity that can be optimized. Fairness and efficiency are crucial in many applications. The ideal solution is both satisfied and optimized, although this may not be possible.
\newline
\textbf{Definition 6 (Efficiency).} The efficiency of an allocation $A=(A_1,A_2,\ldots,A_n)$ is denoted by $e(A)$ and defined as the sum of agent disutilities, i.e., $e(A)=\sum_{i=1}^n V_i(A)$.
\newline
\textbf{Definition 7 (Optimality).} An allocation $A=(A_1,A_2,\ldots,A_n)$ is said to be optimal amongst a set of possible allocation $\mathcal{A}$ if $e(A)={min}_{A^\prime\in \mathcal{A}} e(A^\prime)$.

In this paper, we will consider efficiency optimization given the required extended fairness condition. Specifically, we will address the following question in the section $6$.
\begin{center}
 \emph{What is the computational complexity of computing an optimal allocation while ensuring extended fairness?}   
\end{center}

\section{Relationship Between Fairness Properties}
The division of chores is an example of a fair division problem in which the resource being divided is undesired and each player desires the least amount possible. It is the inverse of the cake-cutting problem, in which the shared resource is desired and each player wants to get the maximum amount. In the general model of cake-cutting with externalities, swap stability guarantees swap envy-freeness and proportionality when the whole chore is allotted. In contrast, using this model for chore division with externalities, we show the relationship among proportionality, swap envy-freeness, and swap stability.

While we omit the externality constraint, the problem of chore division with externalities becomes the standard problem of chore division. In the chore-division problem, we know that envy-free and proportionality are equivalent, while the number of agents is two. In the following proposition, we observe that swap envy-free and proportionality are identical when there are two agents.
\newline
\textbf{Proposition 1.} \textit{Swap envy-freeness and proportionality are equivalent when $n=2$.}

In the general chore division problem, envy-freeness ensures proportionality for an arbitrary number of agents. While taking externalities into account, proportionality doesn't follow from swap envy-freeness for any number $n$ of agents when $n>2$. In \textbf{Example 2}, we ensure swap envy-freeness does not imply proportionality even when $n=3$. Due to space restrictions, the example has been shown in the appendix.

{ While we focus on the case of two agents, we immediately see that proportionality and swap envy-freeness are equivalent. Indeed, proportionality does not ensure swap envy-freeness for an arbitrary number $n$ of agents when $n>2$. Therefore, proportionality does not always imply swap stability, since the notion of swap stability entails swap envy-freeness. \textbf{Example 3} provides a proportional but not swap envy-free allocation even when $n=3$. Due to space restrictions, the example has been shown in the appendix.}

{We have shown swap envy-freeness does not imply proportionality even when the number of agents is three. The definition of swap stability implies swap envy-freeness. The following implication gives a much stronger statement: swap stability implies proportionality for any number of agents.}
\newline
\textbf{Lemma 1.} \textit{ An swap stable allocation entails proportionality.}

As noted, swap stability follows swap envy-freeness in the definition. Lemma $1$ shows that swap stability implies proportionality. In contrast, in the following example, we show that proportionality and swap envy-freeness are not sufficient conditions to guarantee swap stability.
\newline
\textbf{Example 4.} \textit{Suppose there are three agents. Consider the disutility functions: $v_{1,1}(x)=\frac{7}{12},\forall x\in [\frac{1}{3},1]$; $v_{1,2}(x)=\frac{7}{12}, \forall x \in [0,\frac{1}{3}]$ \textit{and} $\frac{1}{3}, \forall x \in [\frac{1}{3},\frac{2}{3}]$; $v_{1,3}(x)=\frac{7}{12}, \forall x \in [0,\frac{1}{3}]$ \textit{and} $\frac{1}{3}, \forall x \in [\frac{2}{3},1]$; $v_{2,1}(x)=v_{2,2}(x)=v_{2,3}(x)=\frac{1}{3},\forall x\in [0,1]$; $v_{3,1}(x)=v_{3,2}(x)=v_{3,3}(x)=\frac{1}{3},\forall x\in [0,1]$. Consider the allocation $A=(A_1,A_2,A_3)$ where $A_1=[0,\frac{1}{3}]$, $A_2=[\frac{1}{3},\frac{2}{3}]$, and $A_3=[\frac{2}{3},1]$. Each agent receives a value less than or equal to $\frac{1}{3}$ and no agent can minimize their utility by swapping his allocation with that of another agent. So the allocation $A=(A_1,A_2,A_3)$ is proportional as well as swap envy-free but not swap stable, since agent $1$ would like to swap the pieces between agents $2$ and $3$, which minimize his utility to $0$ (compared to $\frac{2}{9}$ under $A$).}
\section{Existence of Fair Allocations}
In the classical model, when there are two agents, an envy-free (and therefore proportional) allocation is possible if the chore is cut into two pieces that one agent values equally and the other agent chooses its favorite piece. In this section, we focus on the number of cuts required while finding allocations under fairness constraints. Focusing on the number of cuts is important because agents may prefer to take a contiguous piece rather than a union of crumbs.

Now we show that at most one cut is required to get a proportional and swap envy-free allocation for chore division with externalities.
Theorem $1$ leads to the conclusion.
\newline
\textbf{Theorem 1.} \textit{Let $n = 2$. Then there is an allocation that is proportional, swap envy-free, and therefore needs at most one cut.}

In the general chore division model, envy-free (and hence proportional) allocations that require only $n-1$ cuts are guaranteed to exist. Of course, at least that
many cuts are required because each agent must receive a
piece. In contrast, we show there are some instances where zero cuts are needed to achieve swap stability. To see this, consider an instance where $v_{i,1}(x)=0, \forall x\in [0,1], \forall i \in N$ and the others disutility functions are $v_{i,j}(x)=\frac{1}{n-1},\forall x\in [0,1]$ and allocate the entire cake to agent $1$. 

Now we show that a proportional and swap envy-free allocation can require strictly more than $n-1$ cuts. This lower bound also applies to swap stability, which entails both proportionality and swap envy-freeness.
\newline
\textbf{Theorem 2.} \textit{There may be strictly more than $n-1$ cuts necessary for a proportional and swap envy-free allocation.}

In addition to demonstrating the existence of swap stability under weak assumptions, the following theorem provides an upper limit on the number of cuts required.
\newline
\textbf{Theorem 3.} \textit{ Suppose that the disutility functions are continuous. Then a swap stable is assured to exist and needs $(n-1)n^2$ cuts\cite{Brnzei2013ExternalitiesIC}.}

In the following passage, we show that we are able to find fair allocations under restricted disutility functions. We consider several classes of structured, concisely representable valuations that were originally proposed by Chen et al.\cite{chen2013truth}, and further studied in several recent papers \cite{Caragiannis2011TowardsME,Cohler2011OptimalEC,bei2012optimal,brams2012maxsum}.

\subsection*{Piecewise Constant Disutilities}
A disutility function is considered piecewise constant when the corresponding disutility density function acts as a piecewise constant. This means that the function may be divided into a limited number of subintervals, where the density remains constant within each interval.
Let the chore be given as a set of subintervals $\mathcal{J}=(I_1,I_2,\ldots,I_m)$ such that for all $i,j\in N$, the disutility of agent $i$ corresponding to agent $j$ in the subinterval $I_k$ is given by a disutility function constant on $I_k:$ $v_{i,j}(x)=c_{i,j,k}$, $\forall x \in I_k$ where $I_k=[x_{k-1},x_k]$ and $0=x_0<x_1<\ldots<x_m=1$.

Now we move on to determining an allocation, where we allocate each subinterval in such a way that each agent gets an identically valued allocation. In the following definition, we define the allocation.
\newline
\textbf{Definition 8} \textbf{(Uniform Allocation)}. \textit{An  allocation is regarded as uniform if each agent $i$ receives a contiguous piece $I_{k,i}$ of each subinterval $I_k$ of length $|I_k|/n$ in the context of a chore division problem with piecewise constant disutilities over a set of subintervals $\mathcal{J}=\{I_1,I_2,\ldots,I_m\}$ where $I_{k,i}=[x_{k-1}+(i-1)\alpha_k,x_{k-1}+i\alpha_k]$ and $\alpha_k=|I_k|/n$.}
\newline
\textbf{Theorem 4.} \textit{Consider a chore division instance with externalities, where the disutility functions are piecewise constant. Then the uniform allocation is swap stable and  $mn-1$ cuts are needed.}
\subsection*{Piecewise Linear Disutilities}
Piecewise linear disutility functions encompass a broader range of disutility functions, including the piecewise constant disutility function. An agent's disutility function is considered piecewise linear when its disutility density function is piecewise linear. Piecewise linear functions provide more versatility while maintaining simple representation. The agent's disutility functions may be determined by dividing the interval [0,1] into a series of subintervals, where the agent's value density function has a consistent slope. The agent proceeds to define the limits of each of these subintervals, together with the gradient and intercept of the density function within the subinterval. Let the chore be given as a set of subintervals $\mathcal{J}=(I_1,I_2,\ldots,I_m)$ such that for all $i,j\in N$, the disutility of agent $i$ corresponding to agent $j$ in the subinterval $I_k$ is given by a disutility function linear on $I_k:$ $v_{i,j}(x)=a_{i,j,k}+xb_{i,j,k}$, $\forall x \in I_k$ where $I_k=[x_{k-1},x_k]$ and $0=x_0<x_1<\ldots<x_m=1$.

In the following definition, we define an allocation as the identically valued allocation of piecewise linear functions. 
\newline
\textbf{Definition 9} \textbf{(Sandwich Allocation)}. \textit{The sandwich allocation of a set of subintervals $\mathcal{J}=\{I_1,I_2,\ldots,I_m\}$ is the allocation such that each agent $i$ receives a piece of cake $X_{ij}$ from every subinterval $I_j$ and $X_{ij}$ is defined as:
\begin{equation*}
    X_{ij}=[a_j+(i-1)\alpha_j,a_j+i\alpha_j]\cup [b_j-i\alpha_j,b_j-(i-1)\alpha_j]
\end{equation*}
where $I_j=[a_j,b_j]=[x_{j-1},x_j]$, $\alpha_j=\frac{b_j-a_j}{2n}$.}
\newline
\textbf{Theorem 5.} \textit{Consider a chore division instance with externalities, where the disutility functions are piecewise linear. Then the sandwich allocation is swap stable and $2(n-1)m$ cuts are required.} 

We use the following well-known property of the piecewise linear function, which shows that sandwich allocation gives an identically valued allocation \cite{brams2012maxsum,kurokawa2013cut,chen2013truth}.
\newline
\textbf{Lemma 2.} \textit{Assume that the interval $[a,b]$ is partitioned into $2n$ equal pieces and that an agent has a linear value density on this interval. Let the piece constructed by joining the $i$-th piece from the left (going right) with the $i$-th piece from the right (moving left) be denoted by $X_i$ for $i\in [n]$. That is, $X_1$ is the leftmost and rightmost piece, $X_2$ is the second from the left combined with the second from the right, etc. Then the agent is indifferent between the $X_i$.}
\section{Complexity Considerations}
Branzei et al. \cite{Brnzei2013ExternalitiesIC} extend the Robertson-Webb query model \cite{Cake} to include the following types of queries involving externalities:
\begin{enumerate}
    \item $\textbf{Evaluate}_{i,j}(x,y)$: Output $V_{i,j}([x,y])$.

    \item $\textbf{Cut}_{i,j}(x,\alpha)$: Output $y$ such that $V_{i,j}([x,y])=\alpha$.
\end{enumerate}
The work of Branzei et al.\cite{Brnzei2013ExternalitiesIC} shows that the generalized Robertson-Webb communication model can not provide a proportional protocol, even when there are two agents. This assumption remains true even in the case of chores with externalities.
\newline
\textbf{Theorem 6.} \textit{ In the extended Robertson-Webb model, there is no finite protocol that can determine a proportional allocation of the entire chore, even for two agents \cite{Brnzei2013ExternalitiesIC}.}

{Due to Theorem $6$ and Proposition $1$, we conclude that even with two agents, the generalized Robertson-Webb communication model cannot provide a swap envy-free protocol.}

\section{Optimal Allocation}
The disutility function family that we explore is piecewise constant disutilities. For a disutility function to be piecewise constant, the associated disutility density function must be constant on each subinterval of the chore. There are two main reasons why we are interested in piecewise constant disutility functions. As an example, consider a firm that supervises the operation of the computer network during a conference. Each of its employees has to choose one of the possible shifts, but shifts scheduled at different times of the conference may incur different opportunity costs for various people. Moreover, we can assume that everybody prefers to have just one uninterrupted period to spend at work. Second, the results will be leveraged to address general disutility functions.

This section presents a straightforward technique that can effectively determine an optimal allocation that is both proportional and swap envy-free, given that agents have piecewise constant disutility functions.
In order to discuss computational complexity, we must describe the format of the input to the algorithm. While general disutility functions do not admit concise descriptions, piecewise constant functions have a conveniently simple representation. Each agent reports the boundaries of the subintervals on which the agent’s density function is constant, along with the value of the density function on each of these subintervals.
The size of the input is the number of bits required to report these parameters. We assume that the boundaries of the subintervals and the value of the density function are $p$-bit rationals. A $p$-bit rational is a rational number of the form $a\slash b$, where each $a$ and $b$ is a $p$-bit integer.

When the given disutility functions of all agents are piecewise constant, then the chore can be expressed as a set of subintervals $\mathcal{J}=(I_1,I_2,\ldots,I_m)$ such that for all $i,j\in N$, the disutility of agent $i$ corresponding to agent $j$ in the subinterval $I_k$ is given by a disutility function constant on $I_k:$ $v_{i,j}(x)=c_{i,j,k}$, $\forall x \in I_k$ where $I_k=[x_{k-1},x_k]$ and $0=x_0<x_1<\ldots<x_m=1$. 
\newline
\textbf{Proposition 2.} \textit{Consider a chore division instance $\mathcal{J}$, where the disutility functions are piecewise constants. Then optimal allocation requires at most $m-1$ cuts and can be computed in $mn^2$ queries, where $m$ is the number of intervals in the representation.}

Now we give an algorithm that returns an optimal proportional and swap envy-free allocation in polynomial time in $m$ and $n$, where $m$ is the number of subintervals in the representation. The main idea we use is the conclusion that an algorithm for linear programming executes in polynomial time if all coefficients are $O(p)$-bit rational \cite{Karmarkar1984ANP}. We have the following result:
\newline
\textbf{Theorem 7.} \textit{Assume that $n$ agents have piecewise constant disutilities over a set of intervals $\mathcal{J}=\{I_1,I_2,\ldots,I_m\}$. Then Algorithm \ref{alg2} produces an optimal swap envy-free and proportional allocation in polynomial time.} 
\begin{algorithm}
 \caption{}\label{alg2}
\begin{algorithmic}[1]
\State Let $\mathcal{J}=\{I_1,I_2,\ldots,I_m\}$ be the set of subintervals formed the chore $[0,1]$.
\State  Solve the following linear program:

\begin{align}\label{eq1}
\min \quad
& \sum_{i,j =1}^{n} \sum_{k=1}^m x_{j,I_k} V_{i,j}(I_k)
\end{align}
s.t.
\begin{align}
\sum_{i=1}^n x_{i,I_k}&
= 1 && \forall k\in [m] \label{eq:2} \\
x_{i,k} &\geq 0&& \forall i\in N, \forall k\in [m] \label{eq:3}\\
\sum_{k=1}^m \sum_{j=1}^n x_{j,I_k}V_{i,j}(I_k)&\leq \frac{1}{n} && \forall i\in N \label{eq:4}
\end{align}
 \begin{equation} \label{eq:5}
\begin{split}
\sum_{k=1}^m x_{i,I_k}V_{i,i}(I_k)+x_{j,I_k} V_{i,j}(I_k) & \leq\sum_{k=1}^m x_{j,I_k}V_{i,i}(I_k)+x_{i,I_k} V_{i,j}(I_k)\\ & \forall i,j \in N
\end{split}
\end{equation}
\State Return an allocation which for all $i\in N$ and $I_k\in \mathcal{J}$ allocates an $x_{i,k}$ fraction of subinterval $I_k$ to agent $i$.
\end{algorithmic}
\end{algorithm}

Now we deal with general disutility functions and an arbitrary number of agents. A method is designed to calculate optimal allocations under fairness constraints by approximating disutility functions using piecewise constant functions. This method has a polynomial time-based complexity of $\frac{1}{\epsilon}$, where $\epsilon$ represents the deviation from optimality and the permissible level of envy.
\newline 
\textbf{Lemma 3.} \textit{Given $\epsilon >0$ and general disutility functions $v_{i,j}$. Assume that $v_{i,j}^\prime$ are piecewise constant functions such that for all  $i,j\in N$,
\begin{equation}\label{eq:6}
    v_{i,j}(x)-\frac{\epsilon}{n}\leq v_{i,j}^\prime (x) \leq v_{i,j}(x).
\end{equation}
Let $A=(A_1,A_2,.....,A_n)$ and $A^\prime=(A_1^\prime,A_2^\prime,\ldots,A_n^\prime)$ be optimal proportional allocations with respect to the disutilities $V_{i,j}$ (induced by $v_{i,j}$) and $V_{i,j}^\prime$ (induced by $v_{i,j}^\prime$), respectively. Then $A^\prime$ is an $\frac{\epsilon}{n}$- proportional with respect to $V_{i,j}$ and $e(A^\prime)\leq e(A)+\epsilon$.}
\newline
\textbf{Lemma 4.} \textit{Given $\epsilon >0$ and general disutility functions $v_{i,j}$. Assume that $v_{i,j}^\prime$ are piecewise constant functions such that for all $i,j\in N$,
\begin{equation}\label{eq:7}
    v_{i,j}(x)-\frac{\epsilon}{4}\leq v_{i,j}^\prime (x) \leq v_{i,j}(x).
\end{equation}
Let $A=(A_1,A_2,\ldots,A_n)$ be an optimal proportional and swap envy-free allocation with respect to the disutilities $V_{i,j}$ (induced $v_{i,j}$), and let $A^\prime=(A_1^\prime,A_2^\prime,\ldots,A_n^\prime)$ be an optimal proportional and $\frac{\epsilon}{2}$-swap envy-free allocation with respect to $V_{i,j}^\prime$ (induced by $v_{i,j}^\prime$), respectively. Then $A^\prime$ is an $\frac{\epsilon}{4}$-proportional and $\epsilon$-swap envy-free with respect to $V_{i,j}$ and
$e(A^\prime)\leq e(A)+\frac{n \epsilon}{4}$.}
\newline
\textbf{Lemma 5.}  \textit{Given $\epsilon >0$ and general disutility functions $v_{i,j}$. Assume that $v_{i,j}^\prime$ are piecewise constant functions such that for all $i,j\in N$,
\begin{equation}\label{eq:8}
    v_{i,j}(x)-\frac{\epsilon}{2}\leq v_{i,j}^\prime (x) \leq v_{i,j}(x).
\end{equation}
Let $A=(A_1,A_2,\ldots,A_n)$ be an optimal swap envy-free allocation with respect to the disutilities $V_{i,j}$ (induced $v_{i,j}$), and let $A^\prime=(A_1^\prime,A_2^\prime,\ldots,A_n^\prime)$ be an optimal swap envy-free allocation with respect to $V_{i,j}^\prime$ (induced by $v_{i,j}^\prime$), respectively. Then $A^\prime$ is an $\epsilon$-swap envy-free with respect to $V_{i,j}$ and
$e(A^\prime)\leq e(A)+\frac{n\epsilon}{2}$.}
\newline
\textbf{Theorem 8.} \textit{
Suppose that there are $n$ agents whose disutility functions $v_{i,j}$ are $K$-Lipschitz with $M\leq v_{i,j}(x)$ for some $M\in \mathbb{N}$ , all $i,j\in N$. For any  $\epsilon>0$, there is an algorithm that runs in time polynomial in $n,\log M, K, \frac{1}{\epsilon}$
and computes an $\frac{\epsilon}{n}$-proportional allocation whose efficiency is within $\epsilon$ of the optimal proportional allocation.}
\newline
\textbf{Theorem 9.} \textit{
    Suppose that there are $n$ agents whose disutility functions $v_{i,j}$ are $K$-Lipschitz with $M\leq v_{i,j}(x)$ for some $M\in \mathbb{N}$ , all $i,j\in N$. For any  $\epsilon>0$, there is an algorithm that runs in time polynomial in $n,\log M, K, \frac{1}{\epsilon}$
and computes an $\frac{\epsilon}{4}$-proportional and $\epsilon$-swap envy-free allocation whose efficiency is within $\frac{n \epsilon}{4}$ of the optimal proportional and swap envy-free allocation.}
\newline
\textbf{Theorem 10.} \textit{
     Suppose that there are $n$ agents whose disutility functions $v_{i,j}$ are $K$-Lipschitz with $M\leq v_{i,j}(x)$ for some $M\in \mathbb{N}$ , all $i,j\in N$. For any  $\epsilon>0$, there is an algorithm that runs in time polynomial in $n,\log M, K, \frac{1}{\epsilon}$
and computes an $\epsilon$-swap envy-free allocation whose efficiency is within $\frac{n \epsilon}{2}$ of the optimal proportional and swap envy-free allocation.}

%
%
 \bibliographystyle{splncs04}
 \bibliography{mybibliography}

\begin{thebibliography}{10}
\providecommand{\url}[1]{\texttt{#1}}
\providecommand{\urlprefix}{URL }
\providecommand{\doi}[1]{https://doi.org/#1}

\bibitem{Alon2012SequentialVW}
Alon, N., Babaioff, M., Karidi, R., Lavi, R., Tennenholtz, M.: Sequential voting with externalities: herding in social networks. In: ACM Conference on Economics and Computation (2012)

\bibitem{Arunachaleswaran2019FairAE}
Arunachaleswaran, E.R., Barman, S., Kumar, R., Rathi, N.: Fair and efficient cake division with connected pieces. In: Workshop on Internet and Network Economics (2019)

\bibitem{feeiciencyoffairdivisionwithconnectedpiece}
Aumann, Y., Dombb, Y.: The efficiency of fair division with connected pieces. ACM Transactions on Economics and Computation (TEAC)  \textbf{3},  1 -- 16 (2010)

\bibitem{aumann2012computing}
Aumann, Y., Dombb, Y., Hassidim, A.: Computing socially-efficient cake divisions. In: Adaptive Agents and Multi-Agent Systems. pp. 343--350. International Foundation for Autonomous Agents and Multiagent Systems, Richland, SC (2012)

\bibitem{Aumann2012ComputingSC}
Aumann, Y., Dombb, Y., Hassidim, A.: Computing socially-efficient cake divisions. In: Adaptive Agents and Multi-Agent Systems (2012)

\bibitem{Ayres1969ProductionCA}
Ayres, R.U., Kneese, A.V.: Production, consumption, and externalities. The American Economic Review  \textbf{59},  282--297 (1969)

\bibitem{ADiscreteandBoundedEnvy-FreeCakeCuttingProtocolforAnyNumberofAgents}
Aziz, H., Mackenzie, S.: A discrete and bounded envy-free cake cutting protocol for any number of agents. 2016 IEEE 57th Annual Symposium on Foundations of Computer Science (FOCS) pp. 416--427 (2016)

\bibitem{aziz2021fairness}
Aziz, H., Suksompong, W., Sun, Z., Walsh, T.: Fairness concepts for indivisible items with externalities. arXiv preprint arXiv:2110.09066  (2021)

\bibitem{bei2012optimal}
Bei, X., Chen, N., Hua, X., Tao, B., Yang, E.: Optimal proportional cake cutting with connected pieces. In: Proceedings of the AAAI Conference on Artificial Intelligence. pp. 1263--1269. {AAAI} Press, Toronto, Ontario, Canada (2012)

\bibitem{Bei2012OptimalPC}
Bei, X., Chen, N., Hua, X., Tao, B., Yang, E.: Optimal proportional cake cutting with connected pieces. Proceedings of the AAAI Conference on Artificial Intelligence  (2012)

\bibitem{brams2012maxsum}
Brams, S., Feldman, M., Lai, J., Morgenstern, J., Procaccia, A.: On maxsum fair cake divisions. In: Proceedings of the AAAI Conference on Artificial Intelligence. pp. 1285--1291 (2012)

\bibitem{ANENVY-FREECAKEDIVISIONPROTOCOL}
Brams, S.J., Taylor, A.D.: An envy-free cake division protocol. American Mathematical Monthly  \textbf{102},  9--18 (1995)

\bibitem{brams1996fair}
Brams, S.J., Taylor, A.D.: Fair Division: From cake-cutting to dispute resolution. Cambridge University Press (1996)

\bibitem{brandt2016handbook}
Brandt, F., Conitzer, V., Endriss, U., Lang, J., Procaccia, A.D.: Handbook of computational social choice. Cambridge University Press (2016)

\bibitem{Brnzei2013MatchingsWE}
Br{\^a}nzei, S., Michalak, T.P., Rahwan, T., Larson, K., Jennings, N.R.: Matchings with externalities and attitudes. In: Adaptive Agents and Multi-Agent Systems (2013)

\bibitem{TheQuerycomplexityofcakecutting}
Br{\^a}nzei, S., Nisan, N.: The query complexity of cake cutting. Advances in Neural Information Processing Systems  \textbf{35},  37905--37919 (2022)

\bibitem{Brnzei2013ExternalitiesIC}
Br{\^a}nzei, S., Procaccia, A.D., Zhang, J.: Externalities in cake cutting. In: International Joint Conference on Artificial Intelligence (2013)

\bibitem{AlgorithmsforCdivisionofchores}
Br{\^a}nzei, S., Sandomirskiy, F.: Algorithms for competitive division of chores. Mathematics of Operations Research  (2023)

\bibitem{TheEfficiencyoffairdivision}
Caragiannis, I., Kaklamanis, C., Kanellopoulos, P., Kyropoulou, M.: The efficiency of fair division. Theory of Computing Systems  \textbf{50},  589--610 (2009)

\bibitem{caragiannis2012efficiency}
Caragiannis, I., Kaklamanis, C., Kanellopoulos, P., Kyropoulou, M.: The efficiency of fair division. Theory of Computing Systems  \textbf{50},  589--610 (2012)

\bibitem{Caragiannis2011TowardsME}
Caragiannis, I., Lai, J.K., Procaccia, A.D.: Towards more expressive cake cutting. In: International Joint Conference on Artificial Intelligence (2011)

\bibitem{Chaudhury2020DividingBI}
Chaudhury, B.R., Garg, J., McGlaughlin, P.C., Mehta, R.: Dividing bads is harder than dividing goods: On the complexity of fair and efficient division of chores. ArXiv  \textbf{abs/2008.00285} (2020)

\bibitem{chen2013truth}
Chen, Y., Lai, J.K., Parkes, D.C., Procaccia, A.D.: Truth, justice, and cake cutting. Games and Economic Behavior  \textbf{77}(1),  284--297 (2013)

\bibitem{Cohler2011OptimalEC}
Cohler, Y.J., Lai, J.K., Parkes, D.C., Procaccia, A.D.: Optimal envy-free cake cutting. Proceedings of the AAAI Conference on Artificial Intelligence  (2011)

\bibitem{AlgorithmicSolutionsforEnvy-FreeCakeCutting}
Deng, X., Qi, Q., Saberi, A.: Algorithmic solutions for envy-free cake cutting. Oper. Res.  \textbf{60},  1461--1476 (2012)

\bibitem{Howtocutacakefairly}
Dubins, L.E., Spanier, E.H.: How to cut a cake fairly. American Mathematical Monthly  \textbf{68},  1--17 (1961)

\bibitem{easley2010networks}
Easley, D., Kleinberg, J., et~al.: Networks, crowds, and markets: Reasoning about a highly connected world, vol.~1. Cambridge university press Cambridge (2010)

\bibitem{cakecuttingreallyisnotapieceofcake}
Edmonds, J., Pruhs, K.: Cake cutting really is not a piece of cake. In: ACM-SIAM Symposium on Discrete Algorithms (2006)

\bibitem{ANoteOnCakeCutting}
Even, S., Paz, A.: A note on cake cutting. Discret. Appl. Math.  \textbf{7},  285--296 (1984)

\bibitem{complexitychoredivisio}
Farhadi, A., Hajiaghayi, M.T.: On the complexity of chore division. In: International Joint Conference on Artificial Intelligence (2017)

\bibitem{ghodsi2018fair}
Ghodsi, M., Saleh, H., Seddighin, M.: Fair allocation of indivisible items with externalities. arXiv preprint arXiv:1805.06191  (2018)

\bibitem{Haghpanah2011OptimalAW}
Haghpanah, N., Immorlica, N., Mirrokni, V.S., Munagala, K.: Optimal auctions with positive network externalities. In: ACM Conference on Economics and Computation (2011)

\bibitem{DividingConnectedChoresfairly}
Heydrich, S., van Stee, R.: Dividing connected chores fairly. In: Algorithmic Game Theory (2013)

\bibitem{Karmarkar1984ANP}
Karmarkar, N.: A new polynomial-time algorithm for linear programming. Combinatorica  \textbf{4},  373--395 (1984)

\bibitem{Katz1985NetworkEC}
Katz, M.L., Shapiro, C.: Network externalities, competition, and compatibility. The American Economic Review  \textbf{75},  424--440 (1985)

\bibitem{Krysta2010CombinatorialAW}
Krysta, P., Michalak, T.P., Sandholm, T., Wooldridge, M.: Combinatorial auctions with externalities. In: Adaptive Agents and Multi-Agent Systems (2010)

\bibitem{kurokawa2013cut}
Kurokawa, D., Lai, J., Procaccia, A.: How to cut a cake before the party ends. In: Proceedings of the AAAI Conference on Artificial Intelligence. pp. 555--561 (2013)

\bibitem{Michalak2009OnRC}
Michalak, T.P., Rahwan, T., Sroka, J., Dowell, A.J., Wooldridge, M., McBurney, P., Jennings, N.R.: On representing coalitional games with externalities. In: ACM Conference on Economics and Computation (2009)

\bibitem{mishra2022fair}
Mishra, S., Padala, M., Gujar, S.: Fair allocation with special externalities. In: PRICAI 2022: Trends in Artificial Intelligence: 19th Pacific Rim International Conference on Artificial Intelligence, PRICAI 2022, Shanghai, China, November 10--13, 2022, Proceedings, Part I. pp. 3--16. Springer (2022)

\bibitem{moulin2004fair}
Moulin, H.: Fair division and collective welfare. MIT press (2004)

\bibitem{Papadimitriou1979EfficientSF}
Papadimitriou, C.H.: Efficient search for rationals. Inf. Process. Lett.  \textbf{8}, ~1--4 (1979)

\bibitem{peterson1998exact}
Peterson, E., Su, F.E.: Exact procedures for envy-free chore division. preprint  (1998)

\bibitem{cakecuttingnotjustchildplay}
Procaccia, A.D.: Cake cutting: not just child's play. Commun. ACM  \textbf{56},  78--87 (2013)

\bibitem{Cake}
Robertson, J., Webb, W.: Cake-Cutting Algorithms Be Fair If You Can. CRC Press (1998)

\bibitem{seddighin2019externalities}
Seddighin, M., Saleh, H., Ghodsi, M.: Externalities and fairness. In: The World Wide Web Conference. pp. 538--548 (2019)

\bibitem{ChildrenCryingatbirthdayparties}
Thomson, W.: Children crying at birthday parties. why? Economic Theory  \textbf{31},  501--521 (2007)

\bibitem{Velez2016FairnessAE}
Velez, R.A.: Fairness and externalities. Theoretical Economics  \textbf{11},  381--410 (2016)

\end{thebibliography}
 \section*{Appendix}
 For convenience, we mention the statement of theorem, proposition, and lemma. Note that the equations $(\ref{eq:6})$, $(\ref{eq:7})$, and $(\ref{eq:8})$ are equal to the equations $(\ref{eq:15})$, $(\ref{eq:18})$, and $(\ref{eq:20})$ respectively.
 \section*{Relationship Between Fairness Properties}
\textbf{Proposition 1.} Swap envy-freeness and proportionality are equivalent when $n=2$.\newline
\textit{Proof.} Assume that $A=(A_1,A_2)$ be any swap envy-free allocation.

By the definition of swap envy-freeness, we have
\begin{equation}\label{eq:9}
    V_{i,i}(A_i)+V_{i,3-i}(A_{3-i})\leq  V_{i,i}(A_{3-i})+V_{i,3-i}(A_i).
\end{equation}
Due to normalization of disutilities and inequality $(\ref{eq:9})$, we must have 
\begin{equation}\label{eq:10}
    2 \left(V_{i,i}(A_i)+V_{i,3-i}(A_{3-i})\right)\leq 1.
\end{equation}
Therefore, from (\ref{eq:10}) we get $V_i(A)\leq \frac{1}{2}$.
Thus, swap envy-freeness always guarantees the existence of proportionality.

Conversely, suppose that $A=(A_1,A_2)$ be a proportional allocation.

Due to the definition of proportionality, we get
\begin{equation}\label{eq:11}
 V_{i,i}(A_i)+V_{i,3-i}(A_{3-i}) \leq \frac{1}{2}.
\end{equation}
Due to normalization of disutilities and inequality (\ref{eq:11}), we obtain
\begin{equation}\label{eq:12}
    V_{i,i}(A_{3-i})+V_{i,3-i}(A_i) \geq \frac{1}{2}
\end{equation}
Therefore, the inequalities (\ref{eq:11}) and (\ref{eq:12}) ensure the achievement of swap envy-freeness. \qed

\textbf{Example 2.} \textit{Suppose there are three agents. Consider the disutility functions: $v_{1,1}(x)=\frac{9}{20},\forall x\in [\frac{1}{3},1]$; $v_{1,2}(x)=\frac{9}{20}, \forall x \in [0,\frac{1}{3}]$ \textit{and} $\frac{3}{5}, \forall x \in [\frac{1}{3},\frac{2}{3}]$; $v_{1,3}(x)=\frac{9}{20}, \forall x \in [0,\frac{1}{3}]$ \textit{and} $\frac{3}{5}, \forall x \in [\frac{2}{3},1]$; $v_{2,1}(x)=v_{2,2}(x)=v_{2,3}(x)=\frac{1}{3},\forall x\in [0,1]$; $v_{3,1}(x)=v_{3,2}(x)=v_{3,3}(x)=\frac{1}{3},\forall x\in [0,1]$. The allocation $A=(A_1,A_2,A_3)$, where $A_1=[0,\frac{1}{3}]$, $A_2=[\frac{1}{3},\frac{2}{3}]$ and $A_3=[\frac{2}{3},1]$ is swap envy-free but not proportional. Because agent $1$ receives utility $V_1(A)=V_{1,1}(A_1)+V_{1,2}(A_2)+V_{1,3}(A_3)=0*\frac{1}{3}+\frac{3}{5}*\frac{1}{3}+\frac{3}{5}*\frac{1}{3}=\frac{2}{5}>\frac{1}{3}$.}

\textbf{Example 3.} \textit{Suppose there are three agents. Consider the disutility functions: $v_{1,1}(x)=\frac{6}{7},\forall x\in [0,\frac{1}{3}]$ and $v_{1,1}(x)=\frac{1}{2},\forall x\in [\frac{1}{3},1]$; $v_{1,2}(x)=\frac{3}{42},\forall x\in [\frac{1}{3},\frac{2}{3}]$ and $\frac{1}{2}, \forall x \in [\frac{2}{3},1] $; $v_{1,3}(x)=\frac{1}{2},\forall x\in [\frac{1}{3},\frac{2}{3}]$ and $\frac{3}{42},\forall x\in [\frac{2}{3},1]$; $v_{2,1}(x)=v_{2,2}(x)=v_{2,3}(x)=\frac{1}{3}, \forall x\in [0,1]$; $v_{3,1}(x)=v_{3,2}(x)=v_{3,3}(x)=\frac{1}{3}, \forall x\in [0,1]$; The allocation $A=(A_1,A_2,A_3)$, where $A_1=[0,\frac{1}{3}]$, $A_2=[\frac{1}{3},\frac{2}{3}]$, and $A_3=[\frac{2}{3},1]$ is proportional but not swap envy-free, since agent $1$ can minimize his disutility by swapping his piece among any of others.}

\textbf{Lemma 1.} \textit{ An swap stable allocation entails proportionality.}

\textit{Proof.} Suppose that $A=(A_1,A_2,\ldots,A_n)$ be any swap stable allocation.

Due to the definition of swap stability, we have that for any three agents $i,j,k\in N$,
\begin{equation}\label{eq:13}
    V_{i,j}(A_j)+V_{i,k}(A_k)\leq V_{i,j}(A_k)+V_{i,k}(A_j).
\end{equation}
Summing inequality (\ref{eq:13}) over all $k\in N$, we have
\begin{equation*}
   \sum\limits_{k=1}^n V_{i,j}(A_j)+\sum\limits_{k=1}^n V_{i,k}(A_k)\leq \sum\limits_{k=1}^n V_{i,j}(A_k)+\sum\limits_{k=1}^n V_{i,k}(A_j).
\end{equation*}
Because of $V_i(A)=\sum\limits_{k=1}^n V_{i,k}(A_k)$ and $\bigcup_{k=1}^n A_k =[0,1]$, we get from the above inequality
\begin{equation}\label{eq:14}
     n V_{i,j}(A_j)+V_i(A)\leq V_{i,j}([0,1])+\sum\limits_{k=1}^n V_{i,k}(A_j).
\end{equation}
Again summing inequality (\ref{eq:14}) over all $j\in N$, we obtain
\begin{equation*}
    n \sum\limits_{j=1}^n V_{i,j}(A_j)+\sum\limits_{j=1}^n V_i(A) \leq \sum\limits_{j=1}^n V_{i,j}([0,1])+\sum\limits_{j=1}^n \sum\limits_{k=1}^n V_{i,k}(A_j).
\end{equation*}
 Equivalently,
\begin{equation*}
    n V_{i}(A)+n V_i(A) \leq \sum\limits_{j=1}^n V_{i,j}([0,1])+\sum\limits_{k=1}^n \sum\limits_{j=1}^n V_{i,k}(A_j).
\end{equation*}
 Equivalently,
\begin{equation*}
    2n V_{i}(A)\leq \sum\limits_{j=1}^n V_{i,j}([0,1])+\sum\limits_{k=1}^n V_{i,k}([0,1]).
\end{equation*}

Since $\sum\limits_{j\in N} V_{i,j}([0,1])=1$, we obtain
\begin{equation*}
    2n V_{i}(A)\leq \sum\limits_{j=1}^n V_{i,j}([0,1])+\sum\limits_{k=1}^n V_{i,k}([0,1])=1+1=2.
\end{equation*}
Thus, $V_i(A)\leq \frac{1}{n}$ for any $i\in N$. Therefore, the allocation $A=(A_1,A_2,\ldots,A_n)$ is proportional. \qed
\section*{Existence of Fair Allocations}

\textbf{Theorem 1.} \textit{Let $n = 2$. Then there is an allocation that is proportional, swap envy-free, and therefore needs at most one cut.}\newline
\textit{Proof.} The scenario of the proof is to find a point $\Tilde{y}\in [0,1]$ such that $[0,\Tilde{y}]$ and  $[\Tilde{y},1]$ is a proportional and swap envy-free allocation.\newline
Now define a function $\mathcal{F}:[0,1]\rightarrow{\mathbb{R}}$ such that for all $x\in [0,1]$:
\begin{equation*}
    \mathcal{F}(x)=V_{2,1}([0,x])+V_{2,2}([x,1])-V_{2,1}([x,1])-V_{2,2}([0,x]).
\end{equation*}
It is clear that $\mathcal{F}$ is a continuous function and the values of $\mathcal{F}$ at the points $0$ and $1$ are
$\mathcal{F}(0)=V_{2,2}([0,1])-V_{2,1}([0,1])$ and 
$\mathcal{F}(1)=V_{2,1}([0,1])-V_{2,2}([0,1])$ respectively.
Thus, we get $\mathcal{F}(0)+\mathcal{F}(1)=0$. Assuming that $\mathcal{F}(0)\neq 0$. Therefore, due to the intermediate value theorem, there exists a point $\Tilde{y}\in [0,1]$ such that $\mathcal{F}(\Tilde{y})=0$.

Now we show that the allocation in which agent $1$ chooses his most preferred piece among $\{[0,\Tilde{y}],[\Tilde{y},1]\}$ and allocating the remaining piece to agent $2$ is proportional and swap envy-free.
\newline
Suppose that agent $1$ selects a piece from $\{[0,\Tilde{y}],[\Tilde{y},1]\}$ that is, $[\Tilde{y},1]$. Then the corresponding allocation is $A=(A_1,A_2)$ where $A_1=[\Tilde{y},1]$ and $A_2=[0,\Tilde{y}]$.

Since agent $1$ chooses his most preferred piece $[\Tilde{y},1]$, we must have
\begin{equation*}
    V_{1,1}([\Tilde{y},1])+V_{1,2}([0,\Tilde{y}])\leq V_{1,1}([0,\Tilde{y}])+V_{1,2}([\Tilde{y},1]).
\end{equation*}
Thus, agent $1$ is swap envy-free. Now, via contradiction, we show that agent $1$ has a worth below $\frac{1}{2}$. Suppose that $V_1(A)>\frac{1}{2}$. Then we have 
\begin{align*} 
 \frac{1}{2} &< V_1(A) = V_{1,1}([\Tilde{y},1])+V_{1,2}([0,\Tilde{y}]) \\
       &\leq V_{1,1}([0,\Tilde{y}])+V_{1,2}([\Tilde{y},1]).
\end{align*}
From the above inequality, we get 
\begin{align*} 
 1 &<V_{1,1}([\Tilde{y},1])+V_{1,2}([0,\Tilde{y}])+V_{1,1}([0,\Tilde{y}])+V_{1,2}([\Tilde{y},1]) \\
       &= V_{1,1}([0,1])+V_{1,2}([0,1])=1.
\end{align*}
This is a contradiction, thus $V_1(A)\leq \frac{1}{2}$.\newline
We have seen that the allocation $A$ satisfies fairness notions proportionality and swap envy-freeness for agent $1$.\newline
Now we show that the allocation $A$ also satisfies fairness for agent $2$.\newline By the choice of $\Tilde{y}$, $V_{2,1}([\Tilde{y},1])+V_{2,2}([0,\Tilde{y}])=V_{2,1}([0,\Tilde{y}])+V_{2,2}([\Tilde{y},1])$, and so agent $2$ is not swap-envious. Besides,
\begin{align*} 
 2V_2(A) &=2(V_{2,1}([\Tilde{y},1])+V_{2,2}([0,\Tilde{y}]))\\
       &= V_{2,1}([\Tilde{y},1])+V_{2,2}([0,\Tilde{y}])+V_{2,1}([0,\Tilde{y}])+V_{2,2}([\Tilde{y},1])\\
       &=V_{2,1}([0,1])+V_{2,2}([0,1])=1.
\end{align*}
So we get $V_2(A)=\frac{1}{2}$. Thus, the allocation $A$ is proportional, swap envy-free, and therefore only needs a single cut.

 If $\mathcal{F}(0)=\mathcal{F}(1)=0$. That is, $V_{2,1}([0,1])=V_{2,2}([0,1])$. Because of $V_{2,1}([0,1])+V_{2,2}([0,1])=1$, $V_{2,1}([0,1])=V_{2,2}([0,1])=\frac{1}{2}$. Therefore, $A=([0,1],\emptyset)$ is the required allocation that satisfies proportionality and swap envy-freeness if $V_{1,1}([0,1])\leq V_{1,2}([0,1])$. Otherwise, $A=(\emptyset,[0,1])$.\newline
Thus, there exists a proportional and swap envy-free allocation that requires at most one cut.\qed

\textbf{Theorem 2.} \textit{There may be strictly more than $n-1$ cuts necessary for a proportional and swap envy-free allocation.}

\textit{Proof.} Here we consider an instance and then show that a proportional allocation where $n-1$ cuts are used is not swap envy-free. Consider the value density functions:
$v_{i,j}(x)=\frac{1}{n}+\epsilon,\forall x\in [0,1]$ and for all $i\neq j\in N$,
\begin{equation*}
   v_{i,i}=
   \begin{cases}
    0 & \text{if } x \in [\frac{i-1}{n},\frac{i}{n}]\\
    \frac{(\frac{1}{n}+\epsilon-n\epsilon)n}{n-1} & \text{otherwise}
    \end{cases}
\end{equation*}
where $\epsilon(<\frac{1}{n(n-1)})$ is a small positive number and $i\neq 1$.
\begin{equation*}
    v_{1,1}(x)=
    \begin{cases}
        \frac{(\frac{1}{n}+\epsilon-n\epsilon)}{n-1} & \text{if } x\in [0,\frac{1}{n}]\\
        0 & \text{if } x\in [\frac{2}{n},\frac{3}{n}]\\
        \frac{(n^2-n-1)(\frac{1}{n}+\epsilon-n\epsilon)}{(n-1)(n-2)}& \text{otherwise}
    \end{cases}
\end{equation*}
and  $v_{1,j}(x)=\frac{1}{n}+\epsilon,\forall x\in [0,1]$,
where $j=2,3,....,n$.\newline
In our constructed instance, any proportional allocation $A=(A_1,A_2.\ldots,A_n)$ needs at least $n-1$ cuts. Suppose an agent $k$ does not get any piece, i.e., $A_k=\emptyset$. Since for $j\neq k$,  $v_{k,j}(x)=\frac{1}{n}+\epsilon,\forall x\in [0,1]$. Therefore, $V_k(A)=\sum_{j=1}^n V_{k,j}(A_j)=\frac{1}{n}+\epsilon$.
We allocate the piece $[\frac{i-1}{n},\frac{i}{n}]$ to the agent $i$. Thus, the resulting allocation $A=(A_1,A_2,\ldots,A_n)$ is proportional, where $A_i=[\frac{i-1}{n},\frac{i}{n}]$ for all $i\in N$. Nevertheless, the allocation $A$ is not swap envy-free since, due to swap envy-freeness condition for agent $1$, we get
\begin{equation*}
  V_{1,1}(A_1)+ V_{1,3}(A_3)\leq V_{1,1}(A_3)
  + V_{1,3}(A_1)
\end{equation*}
Equivalently,
\begin{equation*}
    \frac{\frac{1}{n}+\epsilon-n\epsilon}{n(n-1)}+ \frac{\frac{1}{n}+\epsilon}{n} \leq 0 +\frac{\frac{1}{n}+\epsilon}{n}
\end{equation*}

This is a contradiction. Therefore, the allocation $A$ with $n-1$ cuts is not swap envy-free but proportional. This concludes that for a proportional and swap envy-free allocation, at least $n$ cuts are required.\qed 
\subsection*{Piecewise Constant Disutilities}
\textbf{Theorem 4.} \textit{Consider a chore division instance with externalities, where the disutility functions are piecewise constant. Then the uniform allocation is swap stable and  $mn-1$ cuts are needed.}
\newline
\textit{Proof.} The uniform allocation is swap-stable since all agents receive identical pieces. That is, $V_{i,j}(A_p)=V_{i,j}(A_q)$ for all $i,j,p,q \in [n]$, where $A=(A_1,A_2,\ldots,A_n)$ is the uniform allocation. Therefore, the required number of cuts is $(n-1)m+(m-1)=mn-1$.  Due to $m$ subintervals, $m-1$ cuts are required, and for each subinterval, $(n-1)$ cuts are required.
\begin{center}
    \begin{tikzpicture}
        \draw [thick,-] (0,0)--(7,0);
        \draw [-] (.7,-.25)--(.7,.25);
        \draw [-] (1.6,-.25)--(1.6,.25);
        \draw[-](2.6,-.25)--(2.6,.25);
        \draw[-](6,-.25)--(6,.25);
        \draw[-](5.1,-.25)--(5.1,.25);
        \node at (0,-.25) {$x_0$};
        \node at (.5,-.25) {$x_1$};
        \node at (1.4,-.25) {$x_2$};
        \node at (2.4,-.25) {$x_3$};
        \node at (4.9,-.25) {$x_{m-2}$};
        \node at (5.8,-.25) {$x_{m-1}$};
        \node at (7,-.25) {$x_m$};
        \node at (3.85,-.25) {$\ldots$};
        \node at (0.35,.25) {$I_1$};
        \node at (1.15,.25) {$I_2$};
        \node at (2.1,.25) {$I_3$};
        \node at (5.55,.25) {$I_{m-1}$};
        \node at (6.5,.25) {$I_m$};
         \node at (3.85,.25) {$\ldots$};
        
    \end{tikzpicture}
    \begin{tikzpicture}
        \draw [thick,-] (0,0)--(7,0);
        \draw [-] (.7,-.25)--(.7,.25);
        \draw [-] (1.4,-.25)--(1.4,.25);
        \draw[-](2.1,-.25)--(2.1,.25);
        \draw[-](6.3,-.25)--(6.3,.25);
        \draw[-](5.6,-.25)--(5.6,.25);
        \node at (0,-.25) {$x_{k-1}$};
        \node at (7,-.25) {$x_k$};
        \node at (3.85,-.25) {$\ldots$};
        \node at (0.35,.25) {$I_{k,1}$};
        \node at (1.1,.25) {$I_{k,2}$};
        \node at (1.7,.25) {$I_{k,3}$};
        \node at (5.9,.25) {$I_{k,{n-1}}$};
        \node at (6.7,.25) {$I_k,n$};
         \node at (3.85,.25) {$\ldots$};
        
    \end{tikzpicture}
    
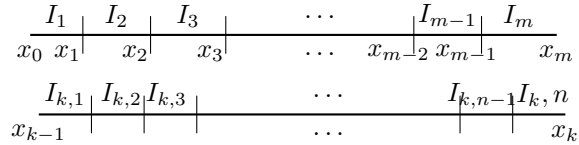
Figure 1: $m-1$ cuts are required due to $m$ subintervals, and for each subinterval $I_k$, $n-1$ cuts are required due to uniform allocation.
\end{center}
\qed
\subsection*{Piecewise Linear Disutilities}
We use the following well-known property of the piecewise linear function, which shows that sandwich allocation gives an identically valued allocation \cite{brams2012maxsum,kurokawa2013cut,chen2013truth}.
\newline
\textbf{Lemma 2.} \textit{Assume that the interval $[a,b]$ is partitioned into $2n$ equal pieces and that an agent has a linear value density on this interval. Let the piece constructed by joining the $i$-th piece from the left (going right) with the $i$-th piece from the right (moving left) be denoted by $X_i$ for $i\in [n]$. That is, $X_1$ is the leftmost and rightmost piece, $X_2$ is the second from the left combined with the second from the right, etc. Then the agent is indifferent between the $X_i$.}
\begin{center}
    \begin{tikzpicture}
        \draw [thick,-] (0,0)--(7,0);
        \draw [thick,-] (0,0)--(0,3);
        \draw [thick,-] (1,2)--(5,3);
        \draw[dashed](1,0)--(1,2);
        \draw[dashed](2,0)--(2,2.25);
        \draw[dashed](3,0)--(3,2.5);
        \draw[dashed](4,0)--(4,2.75);
        \draw[dashed](5,0)--(5,3);
         
        \node at (1,-.25) {$\frac{1}{7}$};
        \node at (2,-.25) {$\frac{2}{7}$};
        \node at (3,-.25) {$\frac{3}{7}$};
        \node at (4,-.25) {$\frac{4}{7}$};
        \node at (5,-.25) {$\frac{5}{7}$};
        \node at (0,-.25) {$0$};
        \node at (7,-.25) {$1$};
        \node at (-.25,2) {$\frac{5}{28}$};
        \node at (-.25,3) {$\frac{27}{140}$};
        \end{tikzpicture}
\end{center}
        
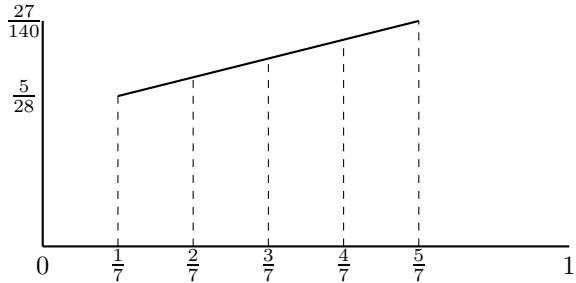
        Figure 2: Linear disutility function $v=\frac{x}{40}+\frac{7}{40}$ and the interval $[a,b]=[\frac{1}{7},\frac{5}{7}]$. Divide the interval into four equal pieces: $[\frac{i-1}{7},\frac{i}{7}]$, where $i=2,3,4,5$. Assume $X_1=[\frac{1}{7},\frac{2}{7}]\cup [\frac{4}{7},\frac{5}{7}]$ and $X_2=[\frac{2}{7},\frac{3}{7}]\cup [\frac{3}{7},\frac{4}{7}]$. Therefore, the disutility for the pieces $X_1$ and $X_2$ are identical and equal to $\frac{13}{245}$.
        
\textbf{Theorem 5.} \textit{Consider a chore division instance with externalities, where the disutility functions are piecewise linear. Then the sandwich allocation is swap stable and $2(n-1)m$ cuts are required.} 

\textit{Proof.} The sandwich allocation is swap stable, as all pieces are the same for every agent due largely to Lemma $2$. That is, $V_{i,j}(A_p)=V_{i,j}(A_q)$ for all $i,j,p,q \in N$ where $A=(A_1,A_2,\ldots,A_n)$ is the sandwich allocation. Moreover, $2(n-1)m$ cuts are needed. Due to each subinterval, $2(n-1)$ cuts are required, and all boundary points of subintervals $x_0,x_1,\ldots,x_n$ lie in the allocated pieces of agent $1$.

\begin{center}
    \begin{tikzpicture}
        \draw [thick,-] (0,0)--(7,0);
        \draw [dashed] (.7,-.25)--(.7,.25);
        \draw [dashed] (1.6,-.25)--(1.6,.25);
        \draw[dashed](2.6,-.25)--(2.6,.25);
        \draw[dashed](6,-.25)--(6,.25);
        \draw[dashed](5.1,-.25)--(5.1,.25);
        \draw [-] (.1,0.25)--(.1,-0.25);
        \draw[-](.6,.2)--(.6,-.2);
        \draw [-] (.85,-.25)--(.85,.25);
        \draw [-] (1.45,-.25)--(1.45,.25);
        \draw [-] (1.7,-.25)--(1.7,.25);
        \draw[-](2.5,-.25)--(2.5,.25);
        \draw[-](2.7,-.25)--(2.7,.25);
        \draw[-](5.9,-.25)--(5.9,.25);
        \draw[-](6.1,-.25)--(6.1,.25);
        \draw[-](5.2,-.25)--(5.2,.25);
        \draw[-](5,-.25)--(5,.25);
        \draw[-](6.9,-.25)--(6.9,.25);
        \node at (0,-.3) {$x_0$};
        \node at (.7,-.3) {$x_1$};
        \node at (1.6,-.3) {$x_2$};
        \node at (2.6,-.3) {$x_3$};
        \node at (5.1,-.3) {$x_{m-2}$};
        \node at (6,-.3) {$x_{m-1}$};
        \node at (7,-.3) {$x_m$};
        \node at (3.85,-.25) {$\ldots$};
        \node at (0.35,.4) {$I_1$};
        \node at (1.15,.4) {$I_2$};
        \node at (2.1,.4) {$I_3$};
        \node at (5.55,.4) {$I_{m-1}$};
        \node at (6.5,.4) {$I_m$};
         \node at (3.85,.4) {$\ldots$};
        
    \end{tikzpicture}
     \begin{tikzpicture}
        \draw [thick,-] (0,0)--(7,0);
        \draw [dashed] (3.5,-.2)--(3.5,.2);
        \draw [-] (.7,-.25)--(.7,.25);
        \draw[-](2.8,-.25)--(2.8,.25);
        \draw[-](4.2,-.25)--(4.2,.25);
        \draw[-](6.3,-.25)--(6.3,.25);
        \node at (-.1,-.25) {$x_{j-1}$};
        \node at (7,-.25) {$x_j$};
        \node at (0.35,.25) {$X_{1j}$};
        \node at (1.75,.25) {$\ldots$};
        \node at (3.5,.25) {$X_{nj}$};
        \node at (5.25,.25) {$\ldots$};
        \node at (6.7,.25) {$X_{1j}$};

    \end{tikzpicture}
\end{center}
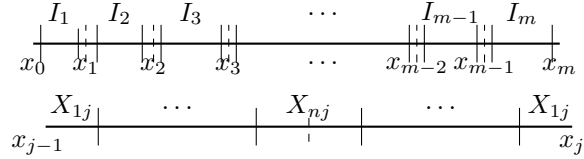
Figure 3: Due to the definition of sandwich allocation, the boundary points of the subintervals lie in the allocation of agent $1$ and for each interval $I_j$, $2(n-1)$ cuts are needed.
\qed
\section*{Optimal Allocation}
When the given disutility functions of all agents are piecewise constant, then the chore can be expressed as a set of subintervals $\mathcal{J}=(I_1,I_2,\ldots,I_m)$ such that for all $i,j\in N$, the disutility of agent $i$ corresponding to agent $j$ in the subinterval $I_k$ is given by a disutility function constant on $I_k:$ $v_{i,j}(x)=c_{i,j,k}$, $\forall x \in I_k$ where $I_k=[x_{k-1},x_k]$ and $0=x_0<x_1<\ldots<x_m=1$. See Figure $4$.

\begin{center}
    \begin{tikzpicture}
        \draw[thick,-](0,0)--(10,0);
        \draw[thick,-](0,0)--(0,4);
         \draw[dashed](2,1)--(4,1);
         \draw[dashed](8,2)--(10,2);
         \draw[dashed](4,2)--(6,2);
         \draw[-](2,.5)--(6,.5);
         \draw[-](6,1.5)--(10,1.5);
         \draw[-](2,0)--(2,3);
         \draw[-](4,0)--(4,3);
         \draw[-](6,0)--(6,3);
         \draw[-](8,0)--(8,3);
         \draw[-](0,1)--(2,1);
         \node at (0,-.25) {$0$};
        \node at (2,-.25) {$\frac{1}{5}$};
        \node at (4,-.25) {$\frac{2}{5}$};
        \node at (6,-.25) {$\frac{3}{5}$};
        \node at (8,-.25) {$\frac{4}{5}$};
        \node at (10,-.25) {$1$};
        \node at (-.25,.5) {$\frac{1}{2}$};
        \node at (-.25,1) {$1$};
        \node at (-.25,1.5) {$\frac{3}{2}$};
        \node at (-.25,2) {$2$};
        \node at (-.25,3) {$3$};
        \node at (-.25,4) {$4$};
        
    \end{tikzpicture}

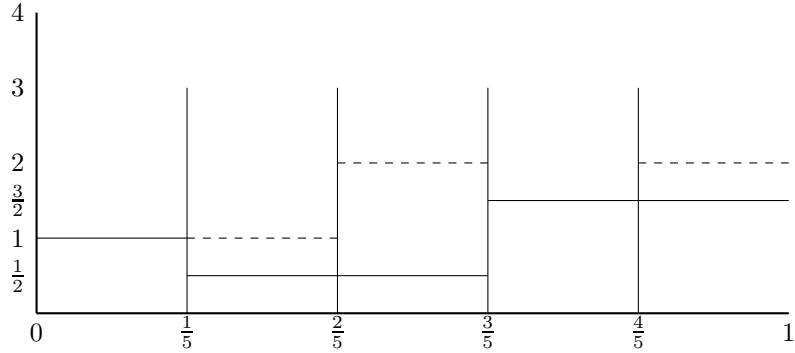
Figure 4: Assume the disutility functions: $v_{1,1}(x)=1$ for $x\in [0,\frac{1}{5}]$, $\frac{1}{2}$ for $x\in [\frac{1}{5},\frac{3}{5}]$, and $\frac{3}{2}$ for $[\frac{3}{5},1]$; $v_{2,1}(x)=1$ for $x\in [\frac{1}{5},\frac{2}{5}]$, $2$ for $x\in [\frac{2}{5},\frac{3}{5}]\cup [\frac{4}{5},1]$;$v_{1,2}(x)=v_{2,2}(x)=0$ for $x\in [0,1]$. The disutility functions are shown using solid and dashed lines. Therefore, the chore is expressed as a set of subintervals $\{I_1,I_2,I_3,I_4,I_5\}$ where $I_i=[\frac{i-1}{5},\frac{i}{5}]$.
\end{center}

\textbf{Proposition 2.} \textit{Consider a chore division instance $\mathcal{J}$, where the disutility functions are piecewise constants. Then optimal allocation requires at most $m-1$ cuts and can be computed in $mn^2$ queries, where $m$ is the number of intervals in the representation.}
\newline
\textit{Proof.} The optimal allocation allocates the interval $I_k$ to the agent $j$, when $\sum_{i=1}^n V_{i,j}(I_k)$ gives the minimum value where $k\in [m]$ and $j\in N$. Thus, the required number of eval queries is $mn^2$ and the allocation needs at most $m-1$ cuts.\qed

\textbf{Description of Algorithm \ref{alg2}}.
The procedure for finding an optimal proportional and swap envy-free allocation is formally given as Algorithm $1$. In Step $1$ of the algorithm, a chore is given as a set of intervals $\mathcal{J}=(I_1,I_2,\ldots,I_m)$ such that for all $i,j\in N$, the disutility of agent $i$ corresponding to agent $j$ in the interval $I_k$ is given by a disutility function constant on $I_k:$ $v_{i,j}(x)=c_{i,j,k}$, $\forall x \in I_k$ where $I_k=[x_{k-1},x_k]$ and $0=x_0<x_1<\ldots<x_m=1$. The linear program (LP) in Step $2$ has variables $x_{iI_{k}}$ for each $i\in N$ and $I_k\in J$ (where $k=1,2,\ldots,m$), which represent the fraction of interval $I_k$ allocated to agent $i$. Since the disutility functions of all agents are constant on each interval $I\in \mathcal{J}$. Hence the disutility of each agent $i\in N$ for a fraction $x_{jI_k}$ of interval $I_k$ allocated to agent $j$ is $x_{jI_k}V_{i,j}(I_k)$, and the agent $i$’s disutility for the piece of agent $j$ is $ \sum_{k=1}^m x_{jI_k}V_{i,j}(I_k)$. So the total disutility of each agent $i$ is $\sum_{j=1}^n \sum_{k=1}^m x_{jI_k}V_{i,j}(I_k)$.  The objective function (\ref{eq1}) then simply gives the efficiency of the allocation. The third constraint (\ref{eq:4}) ensures proportionality, while the fourth constraint (\ref{eq:5}) is simply the swap envy-free constraint.
\newline
\textbf{Theorem 7.} \textit{Assume that $n$ agents have piecewise constant disutilities over a set of intervals $\mathcal{J}=\{I_1,I_2,\ldots,I_m\}$. Then Algorithm \ref{alg2} produces an optimal swap envy-free and proportional allocation in polynomial time.} 
\newline
\textit{Proof.}
The allocation generated is made guaranteed to be both feasible and complete for (\ref{eq:2}) and (\ref{eq:3}) in the algorithm. In addition, (\ref{eq:4}) and (\ref{eq:5}) assure that the allocation is proportional and swap envy-freeness, respectively. Therefore, minimizing efficiency is equated to $(\ref{eq1})$. Because LPs can be solved exactly given rational inputs, it follows that the allocation $(A_1,A_2,\ldots,A_n)$ is optimal amongst allocations that satisfy both proportionality and swap envy-freeness.

It remains to prove that the algorithm runs in time polynomials in the input size. It is sufficient to show that the LP's size is polynomial in input size \cite{Karmarkar1984ANP}. Each $V_{i,j}(I_k)$  is the product of the length of $I_k$ and $c_{i,j,k}$. The first of these factors is an $O(k)$-bit rational, while the second is a $p$-bit rational; hence, the product of these two is an $O(p)$-bit rational. Again, $|\mathcal{J}|=m$. Thus, there are $O(mn)$ variables. As for the number of constraints, $(\ref{eq:2})$ yields $O(m)$, $(\ref{eq:3})$ yields $O(mn)$, $(\ref{eq:4})$ yields $O(n)$, and $(\ref{eq:5})$ yields $O(n^2)$, for a total of $O(mn+n^2)$ constraints. Thus, all of the coefficients are $O(p)$-bit rationals, and there are a polynomial number of variables and constraints.\qed 

\textbf{Lemma 3.} \textit{Given $\epsilon >0$ and general disutility functions $v_{i,j}$. Assume that $v_{i,j}^\prime$ are piecewise constant functions such that for all  $i,j\in N$,
\begin{equation}\label{eq:15}
    v_{i,j}(x)-\frac{\epsilon}{n}\leq v_{i,j}^\prime (x) \leq v_{i,j}(x).           
\end{equation}
Let $A=(A_1,A_2,.....,A_n)$ and $A^\prime=(A_1^\prime,A_2^\prime,\ldots,A_n^\prime)$ be optimal proportional allocations with respect to the disutilities $V_{i,j}$ (induced by $v_{i,j}$) and $V_{i,j}^\prime$ (induced by $v_{i,j}^\prime$), respectively. Then $A^\prime$ is an $\frac{\epsilon}{n}$- proportional with respect to $V_{i,j}$ and $e(A^\prime)\leq e(A)+\epsilon$.}
\newline
\textit{Proof.} To show $A=(A_1,A_2,\ldots,A_n)$ is proportional with respect to $V_{i,j}^\prime$, we have
\begin{equation}\label{eq:16}
   V_{i,j}^\prime(A_j) \leq V_{i,j}(A_j). 
\end{equation}
Summing the inequality $(\ref{eq:16})$ over $j\in N$, we get
\begin{equation}\label{eq:17}
V_i^\prime(A)= \sum_{j=1}^n V_{i,j}^\prime(A_j) \leq V_i(A)=\sum_{j=1}^n V_{i,j}(A_j).
\end{equation}
Due to the fact that $A=(A_1,A_2,\ldots,A_n)$ is proportional with respect to $V_{i,j}$ and from inequality $(\ref{eq:17})$, we obtain
\begin{equation*}
V_i^\prime (A)= \sum_{j=1}^n V_{i,j}^\prime(A_j) \leq V_i(A)=\sum_{j=1}^n V_{i,j}(A_j) \leq \frac{1}{n}. 
\end{equation*}
Thus, we have $V_i^\prime(A)\leq \frac{1}{n}$ for all $i\in N$. So $A=(A_1,A_2,\ldots,A_n)$ is proportional with respect to $V_{i,j}^\prime$.
\newline
Next, we assert that 
\begin{equation*}
\sum_{i}^n V_i^\prime(A^\prime)\leq \sum_{i}^n V_i^\prime(A).  
\end{equation*}
Since $A^\prime=(A_1^\prime,A_2^\prime,\ldots,A_n^\prime)$ is an optimal proportional allocation with respect to $V_{i,j}^\prime$ and $A=(A_1,A_2,\ldots,A_n)$ is a proportional allocation with respect to $V_{i,j}^\prime$. 

From $(\ref{eq:15})$, we must obtain
\begin{equation*}
    \begin{split}
    V_{i}(A^\prime)=\sum_{j=1}^n V_{i,j}(A_j^\prime)&=\sum_{j=1}^n \int_{A_j^\prime} v_{i,j}(x) dx\\&\leq \sum_{j=1}^n \int_{A_j^\prime}(v_{i,j}^\prime(x)+\frac{\epsilon}{n})dx\\&=V_i^\prime(A^\prime)+\frac{\epsilon}{n} \leq \frac{1}{n}+\frac{\epsilon}{n}.
    \end{split}
\end{equation*}
Furthermore, it preserves 
\begin{equation*}
\begin{split}
 \sum_{i=1}^n V_i(A^\prime)=\sum_{i=1}^n \sum_{j=1}^n  V_{i,j}(A_j^\prime)&=\sum_{i=1}^n \sum_{j=1}^n \int_{A_j^\prime} v_{i,j}(x) dx\\
  &\leq \sum_{i=1}^n \sum_{j=1}^n \int_{A_j^\prime}(v_{i,j}^\prime(x)+\frac{\epsilon}{n})dx\\&=\sum_{i=1}^n \sum_{j=1}^n \int_{A_j^\prime}v_{i,j}^\prime(x)dx + \epsilon \\ &\leq\sum_{i=1}^n V_i(A)+\epsilon. 
\end{split}
\end{equation*}

Equivalently, $e(A^\prime)\leq e(A)+\epsilon$.\qed

\textbf{Lemma 4.} \textit{Given $\epsilon >0$ and general disutility functions $v_{i,j}$. Assume that $v_{i,j}^\prime$ are piecewise constant functions such that for all $i,j\in N$,
\begin{equation}\label{eq:18}
    v_{i,j}(x)-\frac{\epsilon}{4}\leq v_{i,j}^\prime (x) \leq v_{i,j}(x). 
\end{equation}
Let $A=(A_1,A_2,\ldots,A_n)$ be an optimal proportional and swap envy-free allocation with respect to the disutilities $V_{i,j}$ (induced $v_{i,j}$), and let $A^\prime=(A_1^\prime,A_2^\prime,\ldots,A_n^\prime)$ be an optimal proportional and $\frac{\epsilon}{2}$-swap envy-free allocation with respect to $V_{i,j}^\prime$ (induced by $v_{i,j}^\prime$), respectively. Then $A^\prime$ is an $\frac{\epsilon}{4}$-proportional and $\epsilon$-swap envy-free with respect to $V_{i,j}$ and
$e(A^\prime)\leq e(A)+\frac{n \epsilon}{4}$.}
\newline
\textit{Proof.} To demonstrate that $A^\prime$ is $\epsilon$-swap envy-free with respect to $V_{i,j}$ follow the lead of
\begin{equation*}
\begin{split}
     V_{i,i}(A_i^\prime)+V_{i,j}(A_j^\prime)&\leq (V_{i,i}^\prime(A_i^\prime)+\frac{\epsilon}{4})+(V_{i,j}^\prime(A_j^\prime)+\frac{\epsilon}{4})\\ &= V_{i,i}^\prime(A_i^\prime)+V_{i,j}^\prime(A_j^\prime )+\frac{\epsilon}{2}\\&\leq (V_{i,i}^\prime(A_j^\prime)+V_{i,j}^\prime(A_i^\prime )+\frac{\epsilon}{2})+\frac{\epsilon}{2}\\&=V_{i,i}^\prime(A_j^\prime)+V_{i,j}^\prime(A_i^\prime )+\epsilon.
\end{split}
\end{equation*}
Since $A^\prime$ is proportional with respect to $V_{i,j}^\prime$, so $A^\prime$ is an $\frac{\epsilon}{4}$-proportional allocation with respect to $V_{i,j}$ followed from $(\ref{eq:18})$. \newline
To show that $A$ is $\frac{\epsilon}{2}$-swap envy-free with respect to $V_{i,j}^\prime$ follow the lead of 

\begin{equation*}
\begin{split}
     V_{i,i}^\prime(A_i)+V_{i,j}^\prime(A_j)&\leq V_{i,i}(A_i)+V_{i,j}(A_j)\\ &\leq V_{i,i}(A_j)+V_{i,j}(A_i)\\&\leq 
(V_{i,i}^\prime(A_j)+\frac{\epsilon}{4})+(V_{i,j}^\prime(A_i)+\frac{\epsilon}{4})\\&=V_{i,i}^\prime(A_j^\prime)+V_{i,j}^\prime(A_i^\prime )+\frac{\epsilon}{2}.
\end{split}
\end{equation*}
Therefore, $A$ is a proportional and $\frac{\epsilon}{2}$-swap envy-free
allocation with respect to $V_{i,j}^\prime$. Proportionality of the allocation with respect to $V_{i,j}^\prime$ followed by Lemma $3$.

Next, we assume that
\begin{equation}\label{eq:19}
 \sum_{i=1}^n V_i^\prime(A^\prime)\leq \sum_{i=1}^n V_i^\prime(A).
\end{equation}
Because of $A^\prime$ is an optimal proportional and $\frac{\epsilon}{2}$-swap envy-free allocation with respect to $V_{i,j}^\prime$, and $A$ is a proportional and $\frac{\epsilon}{2}$-swap envy-free allocation with respect to $V_{i,j}^\prime$.  From $(\ref{eq:18})$, we must have $\sum_{i=1}^n V_{i}^\prime(A)\leq \sum_{i=1}^n V_{i}(A)$.\newline
Furthermore, it yields
\begin{equation*}
\begin{split}
 \sum_{i=1}^n V_i(A^\prime)=\sum_{i=1}^n \sum_{j=1}^n  V_{i,j}(A_j^\prime)&=\sum_{i=1}^n \sum_{j=1}^n \int_{A_j^\prime} v_{i,j}(x) dx\\
  &\leq \sum_{i=1}^n \sum_{j=1}^n \int_{A_j^\prime}(v_{i,j}^\prime(x)+\frac{\epsilon}{4})dx\\&=\sum_{i=1}^n \sum_{j=1}^n \int_{A_j^\prime}v_{i,j}^\prime(x)dx + \frac{n \epsilon}{4} \\ &\leq \sum_{i=1}^n V_i(A)+\frac{n \epsilon}{4}.
\end{split}
\end{equation*}
Equivalently, $e(A^\prime)\leq e(A)+\frac{n \epsilon}{4}$.\qed

\textbf{Lemma 5.}  \textit{Given $\epsilon >0$ and general disutility functions $v_{i,j}$. Assume that $v_{i,j}^\prime$ are piecewise constant functions such that for all $i,j\in N$,
\begin{equation}\label{eq:20}
    v_{i,j}(x)-\frac{\epsilon}{2}\leq v_{i,j}^\prime (x) \leq v_{i,j}(x). 
\end{equation}
Let $A=(A_1,A_2,\ldots,A_n)$ be an optimal swap envy-free allocation with respect to the disutilities $V_{i,j}$ (induced $v_{i,j}$), and let $A^\prime=(A_1^\prime,A_2^\prime,\ldots,A_n^\prime)$ be an optimal swap envy-free allocation with respect to $V_{i,j}^\prime$ (induced by $v_{i,j}^\prime$), respectively. Then $A^\prime$ is an $\epsilon$-swap envy-free with respect to $V_{i,j}$ and
$e(A^\prime)\leq e(A)+\frac{n\epsilon}{2}$.}
\newline
\textit{Proof.} To demonstrate that $A^\prime$ is $\epsilon$-swap envy-free with respect to $V_{i,j}$ follow the lead of
\begin{equation*}
\begin{split}
     V_{i,i}(A_i^\prime)+V_{i,j}(A_j^\prime)&\leq (V_{i,i}^\prime(A_i^\prime)+\frac{\epsilon}{2})+(V_{i,j}^\prime(A_j^\prime)+\frac{\epsilon}{2})\\ &= V_{i,i}^\prime(A_i^\prime)+V_{i,j}^\prime(A_j^\prime )+\epsilon\\&\leq V_{i,i}^\prime(A_j^\prime)+V_{i,j}^\prime(A_i^\prime )+\epsilon\\&
     \leq V_{i,i}(A_j^\prime)+V_{i,j}(A_i^\prime )+\epsilon.
\end{split}
\end{equation*}
The following result implies $A$ is $\epsilon$-swap envy-free with respect to $V_{i,j}^\prime$.
 \begin{equation*}
\begin{split}
     V_{i,i}^\prime(A_i)+V_{i,j}^\prime(A_j)&\leq V_{i,i}(A_i)+V_{i,j}(A_j)\\ &\leq  V_{i,i}(A_j)+V_{i,j}(A_i)\\&\leq (V_{i,i}^\prime(A_j)+\frac{\epsilon}{2})+(V_{i,j}^\prime(A_i)+\frac{\epsilon}{2})
     \\&= V_{i,i}^\prime(A_j)+V_{i,j}^\prime(A_i)+\epsilon.
\end{split}
\end{equation*}
Now, we assert that
\begin{equation}\label{eq:21}
 \sum_{i=1}^n V_i^\prime(A^\prime)\leq \sum_{i=1}^n V_i^\prime(A).
\end{equation}
Because of $A^\prime$ is an optimal swap envy-free allocation with respect $V_{i,j}^\prime$, and $A$ is an $\epsilon$-swap envy-free allocation with respect to $V_{i,j}^\prime$.  From $(\ref{eq:20})$, we must have $\sum_{i=1}^n V_{i}^\prime(A)\leq \sum_{i=1}^n V_{i}(A)$.

Furthermore, it obtains
\begin{equation*}
\begin{split}
 \sum_{i=1}^n V_i(A^\prime)=\sum_{i=1}^n \sum_{j=1}^n  V_{i,j}(A_j^\prime)&=\sum_{i=1}^n \sum_{j=1}^n \int_{A_j^\prime} v_{i,j}(x) dx\\
  &\leq \sum_{i=1}^n \sum_{j=1}^n \int_{A_j^\prime}(v_{i,j}^\prime(x)+\frac{\epsilon}{2})dx\\&=\sum_{i=1}^n \sum_{j=1}^n (\int_{A_j^\prime}v_{i,j}^\prime(x)dx + \frac{|A_j^\prime| \epsilon}{2}) \\ &\leq \sum_{i=1}^n V_i(A)+\frac{ n\epsilon}{2}.
\end{split}
\end{equation*}
Equivalently, $e(A^\prime)\leq e(A)+\frac{ n\epsilon}{2}$.\qed

\textbf{Theorem 8.} \textit{ Suppose that there are $n$ agents whose disutility functions $v_{i,j}$ are $K$-Lipschitz with $M\leq v_{i,j}(x)$ for some $M\in \mathbb{N}$ , all $i,j\in N$. For any  $\epsilon>0$, there is an algorithm that runs in time polynomial in $n,\log M, K, \frac{1}{\epsilon}$
and computes an $\frac{\epsilon}{n}$-proportional allocation whose efficiency is within $\epsilon$ of the optimal proportional allocation.}
\newline
\textit{Proof.} Our approach will depend on an algorithm via lemma reduction. Finding a collection of piecewise constant disutility functions $v_{i,j}^\prime$ that closely resembles the set of real disutility functions $v_{i,j}$ is our first goal.

To find $v_{i,j}^\prime$ as required by Lemma $3$, we know that $v_{i,j}$ are $K$-Lipschitz, i.e., for all $i,j\in N$ and for all $x, y \in [0,1]$,
\begin{equation*}
    |v_{i,j}(x)-v_{i,j}(y)|\leq K.|x-y|.
\end{equation*}
Now, split $[0,1]$ in $\lceil \frac{2nK}{\epsilon} \rceil$ subintervals of size at most $\frac{\epsilon}{2nK}$. Call this new set of subintervals $\mathcal{J}$. For all 
$I_k \in \mathcal{J}$,\newline define
\begin{center}

     $v_{i,j}^* (I_k)=\min\limits_{x\in I_k} v_{i,j}(x)$ \text{and}  $S=\{
 \frac{r}{2^a}: r\in [0,M2^a]\}$,
\end{center}
where $M$ is a lower bound on $v_{i,j}(x)$ for all $i,j\in N$ and $x\in [0,1]$, and $a=\lceil {\log(\frac{2n}{\epsilon})} \rceil $. For each $I_k\in \mathcal{J}$ and $x\in I_k$, let $v_{i,j}^\prime(x)=p_{i,j}^*(I_k)$, where
\begin{equation*}
    p_{i,j}^*(I_k)=\max \{s\in S:s\leq v_{i,j}^*(I_k)\}.
\end{equation*}
The $K$-Lipschitz condition ensures that the disutility function $v_{i,j}$ varies at most $\frac{\epsilon}{2n}$ on each subinterval $I_k$, where $I_k\in \mathcal{J}$, i.e,
\begin{equation*}
    |v_{i,j}(x)-v_{i,j}(y)|\leq \frac{\epsilon}{2n}
\end{equation*}
for any $x,y \in I_k, I_k\in \mathcal{J}$.\newline
Since $a=\lceil {\log(\frac{2n}{\epsilon})} \rceil $, $p_{i,j}^*(I_k)-v_{i,j}^*(I_k)\leq \frac{\epsilon}{2n}$ for each $I_k\in \mathcal{J}$, so $v_{i,j}^\prime$ satisfies the condition $(\ref{eq:15})$. By Lemma $3$, then, an optimal proportional allocation with respect to $v_{i,j}^\prime$ will be  $\frac{\epsilon}{n}$-proportional with respect to the true disutility functions $v_{i,j}$ and have more efficiency. 

Since the $v_{i,j}^\prime$ are piecewise constant and 
 $v_{i,j}^\prime(x)=p_{i,j}^*(I_k)$ for $x\in I_k$ where 
 $k=1,2,\ldots,\lceil \frac{2nK}{\epsilon} \rceil$. Now we consider the set $\mathcal{J}$ of intervals as a chore.
\newline
We can easily find such an allocation by submitting $\mathcal{J}$ to a modified version of Algorithm 1, where the condition $(\ref{eq:5})$ is omitted.

Finally, we specify the running time of the algorithm. First, we have that $|\mathcal{J}|=\lceil {\frac{4K}{\epsilon}}\rceil$. Now the values of $p_{i,j}^*(I_k)$ are $(\log_2 M+a)$-bit rationals \cite{Papadimitriou1979EfficientSF}. We can easily choose the boundaries of the intervals in $\mathcal{J}$ to be $O(\log_2 M+a)$-bit rationals, so that all inputs to Algorithm 1 are $\log_2 M, K$, and $\frac{1}{\epsilon}$. The theorem follows.\qed

\textbf{Theorem 9.} \textit{ Suppose that there are $n$ agents whose disutility functions $v_{i,j}$ are $K$-Lipschitz with $M\leq v_{i,j}(x)$ for some $M\in \mathbb{N}$ , all $i,j\in N$. For any  $\epsilon>0$, there is an algorithm that runs in time polynomial in $n,\log M, K, \frac{1}{\epsilon}$
and computes an $\frac{\epsilon}{4}$-proportional and $\epsilon$-swap envy-free allocation whose efficiency is within $\frac{n \epsilon}{4}$ of the optimal proportional and swap envy-free allocation.}

\textit{Proof.} Split $[0,1]$ in $\lceil\frac{8K}{\epsilon}\rceil$ subintervals of size at most $\frac{8K}{\epsilon}$. So, following from the first portion of the proof of Theorem $8$, we get a collection of piecewise constant disutility functions $v_{i,j}^\prime$ that satisfy condition $(\ref{eq:18})$ where $a=\lceil {2+\log(\frac{2}{\epsilon})} \rceil $. Here $|\mathcal{J}|=\lceil\frac{8K}{\epsilon}\rceil$. 
We can easily find such an allocation by submitting $\mathcal{J}$ to a modified version of Algorithm 1 via Lemma $4$, where the condition $(\ref{eq:5})$ is replaced by
\begin{equation*} 
\begin{split}
\sum_{k=1}^m x_{i,k}V_{i,i}(I_k)+x_{j,k} V_{i,j}(I_k) & \leq\sum_{k=1}^m x_{j,k}V_{i,i}(I_k)+x_{i,k} V_{i,j}(I_k)\\ &+\frac{\epsilon}{2},\forall i,j \in N.
\end{split}
\end{equation*} 
This follows the theorem.\qed

\textbf{Theorem 10.} \textit{ Suppose that there are $n$ agents whose disutility functions $v_{i,j}$ are $K$-Lipschitz with $M\leq v_{i,j}(x)$ for some $M\in \mathbb{N}$ , all $i,j\in N$. For any  $\epsilon>0$, there is an algorithm that runs in time polynomial in $n,\log M, K, \frac{1}{\epsilon}$
and computes an $\epsilon$-swap envy-free allocation whose efficiency is within $\frac{n \epsilon}{2}$ of the optimal proportional and swap envy-free allocation.}

\textit{Proof.} Split $[0,1]$ in $\lceil\frac{4K}{\epsilon}\rceil$ subintervals of size at most $\frac{4K}{\epsilon}$. So, following from the first portion of the proof of Theorem $8$, we get a collection of piecewise constant disutility functions $v_{i,j}^\prime$ that satisfy condition $(\ref{eq:20})$ where $a=\lceil {1+\log(\frac{1}{\epsilon})} \rceil $. Here $|\mathcal{J}|=\lceil\frac{4}{\epsilon}\rceil$. 
We can easily find such an allocation by submitting $\mathcal{J}$ to a modified version of Algorithm 1 via Lemma $5$, where the condition $(\ref{eq:4})$ is omitted.
This follows the theorem.\qed
\section*{Discussion}
In the entire paper, we focus on the study of chore division with externalities using the extended model of the cake-cutting problem with externalities presented by Branzei et al. \cite{Brnzei2013ExternalitiesIC}. We also make the additional assumption that the sum of each agent's multiple valuations should be normalized to 1. In Section $3$, we focus on relationships among the fairness properties, proportionality, swap envy-freeness, and swap stability. Swap stability implies proportionality and swap envy-freeness, but the converse is not true. In Section $4$, we focus on the existence of allocations under fairness constraints. We also focus on the number of cuts required for fair allocations. Focusing on the number of cuts is important because agents may prefer to take a contiguous piece rather than a union of crumbs. We show that a proportional and swap envy-freeness allocation can require at least $n$ cuts. We also demonstrate how to find a swap-stable allocation with finite cuts when the disutility functions are piecewise continuous. In Section $5$, the generalized Robertson and Webb query modal is shown. The existence of a query model and computationally efficient protocols for the computation of swap envy-free and swap stable allocations for any number of agents is one of the significant open problems. In Section $6$, we provide tractable algorithms to find an optimal proportional and swap envy-free allocation under different assumptions regarding the preferences of the agents. Another direction of work is: what happens instead when allowing for both positive and negative values, i.e., when the cake or chore has both desired and undesired pieces?
  
\end{document}